\documentclass[twocolumn]{aastex631}

\usepackage{longtable}
\usepackage{natbib}
\usepackage{amsmath, amsfonts}
\usepackage{amssymb}
\usepackage{gensymb}

\def\simgt{\lower.5ex\hbox{$\; \buildrel > \over \sim \;$}}
\def\simlt{\lower.5ex\hbox{$\; \buildrel < \over \sim \;$}}

\begin{document}

\title{Thermal Properties of 1847 WISE-observed Asteroids}
\author{Denise Hung}
\affiliation{Institute for Astronomy, University of Hawai'i, 2680 Woodlawn Drive, Honolulu, HI 96822, USA}

\author{Josef Hanu\v{s}}
\affiliation{Charles University, Faculty of Mathematics and Physics, Institute of Astronomy, V~Hole{\v s}ovi{\v c}k{\'a}ch 2, 18000 Prague, Czech Republic}

\author{Joseph R. Masiero}
\affiliation{Caltech/IPAC, 1200 E. California Boulevard, MC 100-22, Pasadena, CA 91125, USA}

\author{David J. Tholen}
\affiliation{Institute for Astronomy, University of Hawai'i, 2680 Woodlawn Drive, Honolulu, HI 96822, USA}

\shorttitle{WISE TPM}
\shortauthors{Hung et al.}

\begin{abstract}
We present new thermophysical model (TPM) fits of 1847 asteroids, deriving thermal inertia, diameter, and Bond and visible geometric albedo. We use thermal flux measurements obtained by the Wide-field Infrared Survey Explorer \citep[WISE;][]{Wright10, Mainzer11} during its fully cryogenic phase, when both the 12$\micron$ ($W3$) and 22$\micron$ ($W4$) bands were available. We take shape models and spin information from the Database of Asteroid Models from Inversion Techniques \citep[DAMIT;][]{Durech10} and derive new shape models through lightcurve inversion and combining WISE photometry with existing DAMIT lightcurves. When we limit our sample to the asteroids with the most reliable shape models and thermal flux measurements, we find broadly consistent thermal inertia relations with recent studies. We apply fits to the diameters $D$ (km) and thermal inertia $\Gamma$ (J m$^{-2}$ s$^{-0.5}$ K$^{-1}$) normalized to 1 au with a linear relation of the form $\log[\Gamma]=\alpha+\beta\log[D]$, where we find $\alpha = 2.667 \pm 0.059$ and $\beta = -0.467 \pm 0.044$ for our sample alone and $\alpha = 2.509 \pm 0.017$ and $\beta = -0.352 \pm 0.012$ when combined with other literature estimates. We find little evidence of any correlation between rotation period and thermal inertia, owing to the small number of slow rotators to consider in our sample. While the large uncertainties on the majority of our derived thermal inertia only allow us to identify broad trends between thermal inertia and other physical parameters, we can expect a significant increase in high-quality thermal flux measurements and asteroid shape models with upcoming infrared and wide-field surveys, enabling even more thermophysical modeling of higher precision in the future.
\end{abstract}

\section{Introduction}

Thermal inertia is the measure of a material's resistance to changes in its temperature and can be a useful tool for inferring the surface properties of asteroids. The thermal inertia $\Gamma$ of a material increases with the thermal conductivity $\kappa$, density $\rho$, and specific heat capacity $c$ by $\Gamma = \sqrt{\kappa \rho c}$. Low thermal inertia values are consistent with high porosity or loose regolith, while high values can suggest the presence of conductive materials such as compact rocks and metals. Thermal inertia also governs the Yarkovsky effect, which is a net force that arises from anisotropic thermal emission that perturbs the orbits of asteroids over time \citep[see the reviews by][]{Bottke06, Vokrouhlicky15b}. Because of nonzero thermal inertia, an asteroid will absorb the most heat from the Sun at its subsolar point but only reradiate it some time later in its rotation. The transverse component of the resulting recoil will cause a prograde rotating asteroid to spiral outwards, while a retrograde rotating asteroid will spiral inwards \citep{Bottke06}. The Yarkovsky effect is in part responsible for how asteroid populations are distributed across the solar system today and is important to consider when assessing the impact probabilities of potentially hazardous asteroids \citep[e.g.,][]{Farnocchia13}. 

Asteroid surfaces have been found to have lower than expected thermal inertia values than their material counterparts on Earth, suggesting high porosity. Using a sample of 85 Centaurs and trans-Neptunian objects (TNOs) with thermal observations from Herschel and Spitzer, \citet{Lellouch13} derived thermal inertia that were two to three orders of magnitude smaller than expected for compact ice. Early results from the sample return missions Origins, Spectral Interpretation, Resource Identification, Security, Regolith Explorer (OSIRIS-REx) to (101955) Bennu \citep{Lauretta19} and Hayabusa-2 to (162173) Ryugu \citep{Okada20} have found both asteroids to have thermal inertia between 300 and 350 J m$^{-2}$ s$^{-0.5}$ K$^{-1}$, or much smaller than expected for solid rock. As in situ measurements are presently very few in number, it is unclear whether the subkilometer near-Earth asteroids (NEAs) such as (101955) Bennu and (162173) Ryugu are representative of other B- and C-type asteroids or not. 

It is difficult to comment on the average thermal inertia representative of various asteroid subpopulations as existing thermal inertia estimates only number in the few hundreds and cover a wide range in values, from 0 to nearly 1500 J m$^{-2}$ s$^{-0.5}$ K$^{-1}$ (see Table \ref{tab.prev_gamma}). Previous studies have attempted to identify how thermal inertia might correlate with other physical properties such as spectral type, rotation period, and size \citep[e.g.,][]{Delbo15, Hanus18, Ali-Lagoa20}. Due to the inverse dependency of skin depth with rotation speed, the heat from solar radiation can penetrate the thin topmost regolith layers of slowly rotating asteroids and reach the denser, more highly thermally conductive subsurface material. It was thus hypothesized that slow rotators (i.e., rotation periods longer than 12 hours) should on average present higher thermal inertia than those of fast rotators. While early studies such as \citet{Harris16} and \citet{Marciniak18} appeared to confirm this trend, later studies with larger samples of slow rotators \citep[e.g.,][]{Marciniak19, Ali-Lagoa20, Marciniak21} found no such relation. 

The inverse correlation between asteroid thermal inertia and size was first identified by \citet{Delbo07}. We would expect large asteroids to have low thermal inertia values as their weathered surfaces have accumulated several layers of regolith over hundreds of millions of years. On airless bodies, heat can only be transported through direct conduction or inefficient radiation \citep{Gundlach13}. As the grain size of the regolith decreases (e.g., the finer grains in mature regolith), the area of contact between grains also decreases, reducing the thermal conductivity of the material and thus the thermal inertia. The thermal conductivity of the finely powdered lunar regolith is three orders of magnitude lower than that of compact rock \citep{Delbo15}, corresponding to a low thermal inertia of around 50 J m$^{-2}$ s$^{-0.5}$ K$^{-1}$ \citep[e.g.,][]{Hayne17}.

Simple thermal models such as the widely used Near-Earth Asteroid Thermal Model \citep[NEATM;][]{Harris98} can derive diameters and albedos in the many cases where various physical properties of the asteroids are unknown. However, the thermal inertia can only be constrained using more sophisticated models, such as thermophysical models (TPMs) which numerically solve the heat transfer equation on a large number of plane surface facets to fully represent an asteroid's surface temperature distribution, generating synthetic flux values to compare with observations \citep[e.g.,][]{Spencer89, Lagerros96, Delbo07, Rozitis11}. Whereas simple thermal models assume a spherical asteroid so that they can be widely applied to any thermal data set, TPMs require shape models as input, which are unknown for the vast majority of asteroids. The absence of shape information or thermal data precludes the use of a TPM, and large uncertainties on the derived thermal inertia estimates is common due to the scarcity of high-quality data \citep[see, e.g.,][]{Hanus18}.

The accuracy of physical properties derived with a TPM improves with greater detailed shape models and higher precision multiband infrared data. However, it is not uncommon to see large reduced $\chi^2$ values in TPM fits, due to systematic uncertainties that might be introduced in both thermal observations (e.g., as a result of flux offsets between different instruments) and thermophysical modeling. The recent rise of high-precision thermal infrared surveys of asteroids such as the Wide-field Infrared Survey Explorer \citep[WISE;][]{Wright10} has motivated the need for further optimization in TPMs, as the scale of model uncertainties are now comparable to that of measured flux uncertainties \citep{Delbo15}. The accuracy of shape models is now in some cases a limiting factor, where differences between the assumed shape model and the asteroid's true shape can lead to poor fits \citep[e.g.,][]{Rozitis14}. In addition, current TPMs are largely limited to using convex models with an average albedo and roughness rather than detailed surfaces. However, asteroid shape models have become more common in recent years due to both the public availability of fast, robust lightcurve inversion codes \citep{KaasalainenT01} and a growing archive of photometric data \citep{Warner09}.

The lightcurve inversion code works best with temporally dense photometry, where there are hundreds of measurements over one revolution. However, the code can also be successful with combined independent internally calibrated data sets of temporally sparse photometry, such as measurements obtained by all-sky surveys that only have a few measurements per night. These all-sky surveys provide the largest quantity of photometric data, though unique spin and shape solutions can be difficult to find due to high noise in the data. \citet{Durech19} combined the Lowell Observatory photometric database with the more accurate sparse photometry in Gaia Data Release 2 \citep[DR2;][]{Gaia18}, which were both available for over 5000 asteroids. However, the authors were only able to derive roughly 1000 unique shape models through lightcurve inversion due to limitations imposed by the poor photometric accuracy of the Lowell data as well as the small number of Gaia DR2 data points. The latter issue will resolve itself once the final Gaia catalog is released, and more powerful upcoming all-sky surveys will enable even more accurate spin and shape models. As our knowledge of the physical properties of asteroids increases, the better insight we will have on both individual asteroids and on the population as a whole.

\section{Data and Methodology}

For our TPM, we use the C++ code developed by \citet{Delbo07}, which is publicly available\footnote{\url{https://www.oca.eu/images/LAGRANGE/pages_perso/delbo/thermops.tar.gz}} and one of the most commonly used standard implementations in the field. The TPM takes input information about the asteroid in terms of thermal flux, an ephemeris, a shape model, rotation period, and pole orientation. Our selection criteria for each parameter are chosen largely following the methodology of \citet{Hanus15} and references therein, which we summarize here.

\subsection{Thermophysical Modeling Inputs}\label{sec.tpminputs}

For the flux observations, we use the thermal infrared data taken by the Wide-field Infrared Survey Explorer \citep[WISE;][]{Wright10, Mainzer11} during its fully cryogenic phase between 2010 January 14 to August 6 where both the 12$\micron$ ($W3$) and 22$\micron$ ($W4$) bands were available. Asteroids in this time period were at most observed by WISE in two epochs, with many only having observations in a single epoch. We restrict the TPM application to WISE-observed asteroids with both $W3$ and $W4$ band data. By doing so, we avoid the complications of the presence of reflected sunlight in the 3.4$\micron$ ($W1$) and 4.6$\micron$ ($W2$) bands, which our TPM cannot properly model, though a number of past studies using WISE data with other models have made some use of these bands \citep[e.g.,][]{Ali-Lagoa13, Myhrvold18, Rozitis18, Jiang19, Yu21}. The inclusion of two bands additionally places greater constraints on the TPM fit and size scaling, allowing us to determine properties such as the asteroid's thermal inertia. 

We select which WISE data to use in the TPM largely following the methodology of \citet{Masiero11}. We first obtain R.A./decl./time values from the Minor Planet Center (MPC) observation file\footnote{\url{http://www.minorplanetcenter.net/iau/ECS/MPCAT-OBS/MPCAT-OBS.html}} by selecting all observations submitted from observatory code C51. We then use these ephemerides as input to query the WISE All-Sky Single Exposure (L1b) Source Table through the Gator tool provided by the Infrared Science Archive \citep[IRSA;][]{irsa}.\footnote{\url{https://irsa.ipac.caltech.edu/cgi-bin/Gator/nph-scan?mission=irsa&submit=Select&projshort=WISE}} We limit our search radius to within 2 s and 0.3$\arcsec$ of the time and position of the MPC reported detections in order to minimize the inclusion of unrelated observations. We search for possible star and galaxy contaminants using the positions of the L1b catalog to query the AllWISE Atlas Metadata Table (also served by IRSA) within a search radius of 6.5$\arcsec$, which is equivalent to the $W1$, $W2$, and $W3$ beam sizes. We discard all asteroid observations at positions where a source was found. 

We only include observations with artifact flags of cc\textunderscore flags\,=\,0 or p, which indicate no or possible contamination. The calibration software was found to be overly aggressive in artifact flagging, with cc\textunderscore flags\,=\,0 and cc\textunderscore flags\,=\,p detections having similar fluxes, so the majority of the latter can be treated as reliable \citep{Masiero11}. We use quality flags ph\textunderscore qual\,=\,A, B, or C, which respectively correspond to signal-to-noise (S/N) ratios of S/N$\geq10$, $3<$S/N$<10$, and $2<$S/N$<3$. We impose a constraint of a frame quality score qual\textunderscore frame\,=\,10 to ensure we do not include observations corrupted by image artifacts or electronic noise. We require a South Atlantic Anomaly (SAA) separation value of saa\textunderscore sep\,$>0$, which indicates that WISE was outside of the SAA boundary. Due to concerns regarding nonlinearity and saturation for very bright objects, we exclude observations brighter than $W3 = -2$ and $W4 = -6$ in Vega magnitudes \citep{Cutri12}. 

As a precaution against contamination from spurious sources, we only include observations with $W3$ and $W4$ band errors of $\sigma \leq 0.25$ mag. However, if we have an observation with $\sigma_{W3} \leq 0.25$ mag, we also include the $W4$ observation even if its error is larger than our cutoff. The TPM benefits greatly from the inclusion of multiple bands. The WISE photometric software simultaneously fits a profile to all four bands, using information from each to find the center. The higher S/N observation will control the fit centroid, so we can be confident that the source is present, with the profile scaled to the observed flux. To further ensure that we have two bands to use in our thermophysical modeling, we only include asteroids that have total observations in the $W3$ or $W4$ band of at least 40\% of the other band after our data use cuts, with a minimum of five observations per band. The WISE catalog uncertainties do not include the 0.03 mag systematic internal repeatability component found in comparisons with other catalogs\footnote{\url{https://wise2.ipac.caltech.edu/docs/release/allsky/expsup/sec6_3b.html}} \citep{Cutri12}. To account for this, we add 0.03 mag in quadrature to all reported errors. We convert from magnitudes to flux density with the zero-points in \citet{Wright10} modified for red sources: 306.681 Jy for $W1$, 170.663 Jy for $W2$, 31.3684 Jy for $W3$, and 7.9525 Jy for $W4$. The central wavelengths for $W3$ and $W4$ are shifted to $\lambda_{0, W3}$ = 11.0984 $\mu$m and $\lambda_{0, W4}$ = 22.6405 $\mu$m to correct for the discrepancy observed in photometric calibration tests between blue- and red-spectrum objects. We apply the \citet{Wright10} color corrections for the $W3$ and $W4$ fluxes when passing the thermal data into the TPM.

The asteroid ephemerides used in the TPM are generated such that they span 50 rotations (or a minimum of 10 days) before the first observation to a few days after the last observation in one-day steps. The initial value of the temperature is set at a value given by the first ephemeris point, so setting the start of the ephemerides earlier gives the TPM time to find a more realistic temperature by the time it reaches the first observing epoch. We use the IMCCE Miriade service\footnote{\url{http://vo.imcce.fr/webservices/miriade/}} for its ease of extracting computed ephemerides via HTTP requests. The dates are then corrected for the light travel time before being passed into the TPM.

The TPM uses a convex polyhedron with triangular facets for the shape model. The shape models and spin information we use are sourced from the Database of Asteroid Models from Inversion Techniques \citep[DAMIT;][]{Durech10}\footnote{\url{https://astro.troja.mff.cuni.cz/projects/damit/}}, last retrieved on 2020 November 1. DAMIT also provides its public lightcurve inversion code, which we discuss in \S\ref{sec.shape}. For many of the asteroids, the pole orientation is ambiguous, resulting in multiple (usually two) possible shape models and associated spin information for one asteroid. In these cases, we run the TPM on each shape model.

\subsection{Thermophysical Modeling Application}\label{sec.tpm}

We apply the TPM following the methodology of \citet{Hanus15}. The TPM is run over a grid of three free parameters: the thermal inertia $\Gamma$, Bond albedo $A$, and the surface roughness. The presence of craters on an asteroid will have an effect on its thermal infrared flux due to the interplay between the craters' shadows and heat transfer on the asteroid's surface. In our TPM application, the effects of the surface roughness are described by two parameters, the crater opening angle $\gamma_c$ and crater density $\rho_c$ (Table \ref{tab.roughness}). For a given pair of $\gamma_c$ and $\rho_c$, the TPM computes the \citet{Hapke84} mean surface slope $\bar{\theta}$, which represents the macroscopic roughness angle and is defined by

\begin{equation}
\tan \bar{\theta} = \frac{2}{\pi} \int_{0}^{\pi/2} \tan \theta a(\theta) d\theta
\end{equation}

\noindent where $\theta$ is the angle of a given facet from horizontal and $a(\theta)$ is the distribution of the surface slopes, which are respectively analogous to $\gamma_c$ and $\rho_c$. The thermal emission spectrum depends only on $\bar{\theta}$, for which different combinations of $\gamma_c$ and $\rho_c$ can yield the same value. Any two surfaces with the same value of $\bar{\theta}$ will produce similar disk-integrated thermal emission spectra. The exact values for $\gamma_c$ and $\rho_c$ thus do not matter for the TPM application beyond the effective overall roughness they produce \citep{Emery98}.

We fix the bolometric emissivity $\epsilon$ at 0.9, which is the typical average spectral emissivity for the WISE band wavelengths we are using. The TPM is also run using the \citet{Lagerros96,Lagerros97,Lagerros98} approximation, which uses a simplified heat diffusion model for the craters and is suitable for small phase angle observations, saving significant computational time over the full solution.

\begin{deluxetable*}{cccc}
\tablecolumns{4}
\tablecaption{Surface Roughness\label{tab.roughness}}
\tablehead{
\colhead{Opening Angle ($\degree$)} & \colhead{Crater Density} & \colhead{Hapke Mean Surface Slope ($\degree$)} & \colhead{Qualitative Roughness} \\
\colhead{$\gamma_c$} & \colhead{$\rho_c$} & \colhead{$\bar{\theta}$} & \colhead{}
}
\startdata
0  & 0.0 & 0.0 & None \\
30 & 0.3 & 3.9 & Low \\
40 & 0.7 & 12.6 & Moderate \\
41 & 0.9 & 16.5 & Moderate \\
50 & 0.5 & 12.0 & Moderate \\
60 & 0.9 & 26.7 & High \\
70 & 0.7 & 27.3 & High \\
88 & 0.5 & 35.8 & High \\
89 & 0.7 & 46.8 & Extreme \\
90 & 0.9 & 55.4 & Extreme \\
\enddata
\tablecomments{This table contains the 10 pairs of opening angles and crater densities we use to approximate the surface roughness in the TPM. The relationship between the opening angle and crater density can produce degenerate overall roughness profiles, described by the Hapke mean surface slope, so these pairs were chosen with only the effective roughness in mind.}
\end{deluxetable*}

\begin{figure*}
\centering
\includegraphics[width=2.0\columnwidth]{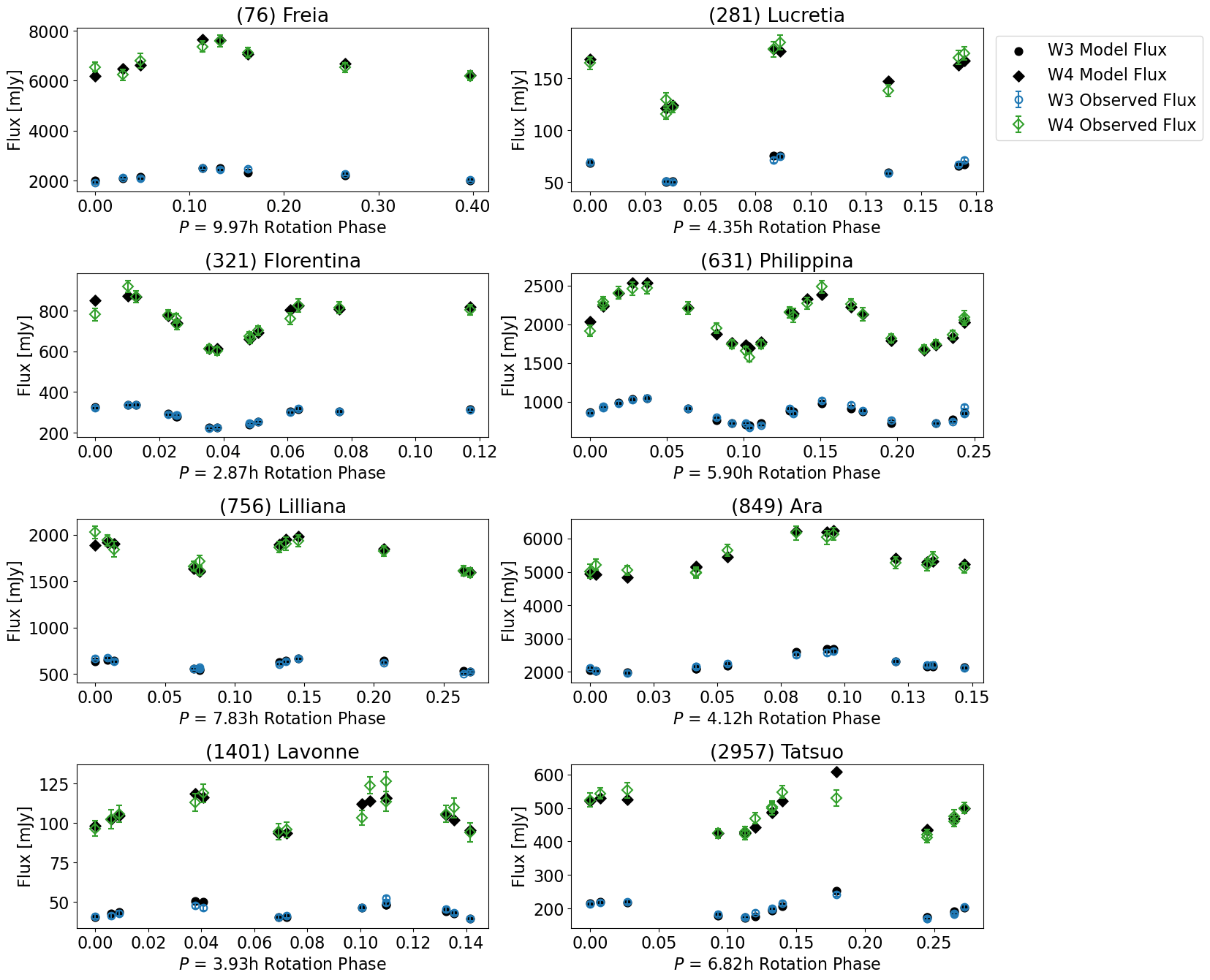} 
\caption{Example TPM fits to WISE thermal data for eight asteroids, folded on the rotation phase. The vast majority of asteroids in our sample were only observed by WISE over one to two days. Due to the cadence and the small number of WISE observations, the rotation phase is usually only sparsely sampled with large gaps in coverage \citep{Wright10}.}
\label{fig.thermal_lc}
\end{figure*}

We run the TPM on 10 loops, one for each surface roughness profile. At the start of each loop, we set the Bond albedo to 0.08 and the thermal inertia to $10^4$ J m$^{-2}$ s$^{-0.5}$ K$^{-1}$ to find an initial diameter estimate. The loop then runs the thermal inertia through progressively lower values down to 0 J m$^{-2}$ s$^{-0.5}$ K$^{-1}$, using progressively finer steps as the TPM fit may become more sensitive to low thermal inertia values. Most asteroids are expected to have significantly lower thermal inertia than our initial value, so we run the thermal inertia from high to low values to give the TPM a few iterations to reach a realistic albedo and size estimate for the asteroid. The TPM scales the surface equivalent size, i.e., the diameter of a sphere with the same surface as the shape model, and from it computes the volume equivalent diameter $D$. The albedo is updated for each iteration based on the previously found diameter. The $H$ and $G$ parameters are used to convert the diameter from one TPM iteration to a more realistic albedo to use on the proceeding iteration. The relation between the diameter $D$ in kilometers and the visible geometric albedo $p_{V}$ \citep{Harris02} is

\begin{equation}\label{eq.d}
D = \frac{1329}{\sqrt p_\mathrm{V}} 10^{-0.2 H}
\end{equation}

The Bond albedo $A$ is then obtained using the phase integral $q$ \citep{Bowell89} with

\begin{equation}\label{eq.albedo}
A \approx q p_\mathrm{V} \approx (0.290 + 0.684 G) p_\mathrm{V}
\end{equation}

The values for the absolute magnitude $H$ and slope $G$ are taken from the Asteroid Absolute Magnitudes and Slopes \citep[AAMS;][]{Muinonen10, Oszkiewicz11} in the Planetary Data System. We do not account for the uncertainty in the $H$ and $G$ values. We note that the thermal inertia has been found to not be strongly dependent on $H$ \citep{Hanus18}. In addition, changes of $\Delta0.5$ mag in $H$ are compensated by a change in $A$ (and therefore $p_{V}$), resulting in a similar $D$ in Equation \ref{eq.d} \citep{Hanus18}. There is no information on $G$ for the majority of objects, and thus it is assumed to be 0.15 for many of the asteroids in our sample.

The TPM calculates a $\chi^2$ value for each iteration with

\begin{equation}\label{eq.tpmchi}
\chi^2 = \sum_i^N\frac{(F_{\mathrm{obs}, i} - F_{\mathrm{mod}, i} s^2 - F_{\mathrm{ref}, i})^2}{\sigma_{\mathrm{obs}, i}^2}
\end{equation}

\noindent where $N$ is the total number of observations over all bands, $F_{\mathrm{obs}}$ is the observed flux, $F_{\mathrm{mod}}$ is the modeled flux, $s$ is the mesh scaling factor, $F_{\mathrm{ref}}$ is the observed reflected flux (assumed to be zero in our application), and $\sigma_{\mathrm{obs}}$ is the uncertainty in the observed flux (Fig. \ref{fig.thermal_lc}).

We caution the reader that this $\chi^2$ value only directly accounts for the uncertainties in the thermal fluxes. The TPM does not account for any uncertainties in either the shape model or the rotation state of the asteroid, so the uncertainties on the derived parameters are best taken to be lower limits. We find the best fit by the TPM solution that yields the smallest $\chi^2$ value. We then estimate the uncertainties in the parameters found by the TPM, the thermal inertia $\Gamma$, diameter $D$, and Bond albedo $A$.

We use an empirical approach to obtain the approximate 1$\sigma$ upper and lower bounds of each parameter, where the error bars span all parameter values corresponding to solutions with $\chi_{\mathrm{red}}^{2} < \chi_{\mathrm{min}}^{2}$ (1 + $\sigma$), where $\sigma$ = $\sqrt{2 \nu}/\nu$ and $\nu$ is the number of degrees of freedom \citep{Press86}. This approach has been commonly used for TPM studies \citep[e.g.,][]{Emery14, Hanus15} and gives reliable uncertainties where $\chi_{\mathrm{red}}^{2} \sim 1$, though this condition is often not met. As we do not know the exact roughness of the asteroid's surface, all solutions here include all roughness profiles to provide more robust uncertainty ranges for each parameter. In the few cases where we see more than one local minimum, we estimate each parameter's uncertainty from the contiguous roughness profile curves containing the best-fit solution.

\subsection{Shape Models}\label{sec.shape}

Shape models are derived using the lightcurve inversion method and codes developed by \citet{KaasalainenT01} and \citet{Kaasalainen01}, which are publicly available on DAMIT. Many of the lightcurves found on DAMIT can be sourced back to the Asteroid Lightcurve Data Exchange Format database\footnote{\url{http://alcdef.org/}} \citep[ALCDEF;][]{Warner11}. We largely follow the methodology of \citet{Hanus11} using the same code. We summarize this process below.

As input, the code takes any number of lightcurves for an asteroid. As with the TPM files, we take the ephemerides from the IMCCE Miriade service, which serve to correct the data for light travel time. In cases where the lightcurves are calibrated, we transform the photometry to intensity, corrected so the asteroid is 1 au from the Sun and 1 au from the observer. If the calibration is not pertinent, such as for the WISE data, we simply normalize the intensity to a mean value of 1. In the case of sparse data, i.e., those from all-sky surveys, we treat the photometry as calibrated. The code does not take into account any flux uncertainties in the lightcurve data.

Obtaining a model first requires finding the correct period for the asteroid. To do this, we make use of a gradient-based algorithm known as convex inversion (CI), which converges to a local minimum given by the initial values of the rotation state parameters: the sidereal rotation period and pole direction. The smallest separation $\Delta P$ between two neighboring local minima in the period spectrum of the lightcurve fit can be approximated by a simple relation:

\begin{equation}
\frac{\Delta P}{P} = \frac{1}{2} \frac{P}{T}
\end{equation}

\noindent where $P$ is the rotation period and $T$ is the temporal coverage of the optical data. After each step, $\Delta P$ is recomputed from the previous period. Using $0.5 \Delta P$ as our period step ensures that we sample all local minima within a given period interval no matter what initial $P$ value is used. For each sample period, we run the CI with 10 different pole orientations isotropically distributed on a sphere. A unique period is found if there are no other solutions with a $\chi_{\mathrm{LC}}^{2}$ below a user-defined threshold multiplied by the $\chi_{\mathrm{LC}}^{2}$ minimum. If the period search is successful, we run the CI with the unique period on a finer grid of pole orientations and look for pole solutions below a new user-defined $\chi_{\mathrm{LC}}^{2}$ cutoff. Due to ambiguities in pole orientation, the search may yield multiple accepted pole solutions, therefore leading to multiple shape models, one for each pole. We identify the pole solution as unique if there are at most two solutions below our defined threshold: the pole and its mirror counterpart. If there is a unique pole solution, we run the CI for each set of unique rotation state parameters and obtain the global shape solution in the representation of a Gaussian image. We then compute the convex polyhedron from the Gaussian image with the Minkowski algorithm, which helps to stabilize the shape solution \citep{Kaasalainen01}. This polyhedron can then be converted to a shape model with triangular facets that is compatible with the TPM code we use.

One of our goals in this paper is to improve existing shape models from the DAMIT database with WISE observations. In order to accomplish this, we derive new shape models by taking the existing lightcurves and appending WISE data onto them. We select the WISE data again in the cryogenic phase of the mission so that we minimize the temporal gap between the optical lightcurve data and thermal data. In addition to the $W3$ and $W4$ bands, we now also include the $W1$ and $W2$ bands for the purpose of the lightcurve inversion in deriving a shape model. The lightcurve inversion method is best suited for optical lightcurves such as the reflected sunlight component in the $W1$ and $W2$ bands. However, the shapes of thermal lightcurves have been found to be similar enough to that of optical lightcurves that we can use the $W3$ and $W4$ data as well for our application \citep{Durech18b}.

The $W1$ and $W2$ data fall under the same constraints we placed on the $W3$ and $W4$ data in \S\ref{sec.tpminputs}, although now we do not explicitly impose a minimum number of observations per band. Each WISE band is a separate lightcurve, so we get a total of four lightcurves out of one observing epoch if data are available in every band. The WISE data we use fall under the same constraints for keeping observations as in \S\ref{sec.tpminputs}, with the only change being that we now only keep observations with errors of $\sigma \leq 0.25$ mag as a precaution against including spurious sources. While the TPM uses multiple bands and can control the fit centroid with a higher S/N data point in one band, the lightcurve inversion treats each band independently. 

The reader might wonder if we lose some information about the thermal properties by treating the thermal data lightcurves (in this case, the $W1$, $W2$, $W3$, and $W4$ bands) as optical. The presence of a nonzero thermal inertia will introduce a phase offset between the thermal and optical lightcurves, which will in turn shift the lightcurve inversion code's best-fit rotation period to compensate. However, a unique period often cannot be determined with sparse data alone. The addition of supplemental data such as the WISE photometry will better constrain the rotation state and the shape, particularly for shape solutions that were based only on sparse data or very few dense optical lightcurves. The period is likely to be better constrained by both the abundance of lightcurve coverage around the WISE photometry as well as the fits to the primarily reflected light $W1$ and $W2$ data points. Simultaneously applying the inversion to both optical and thermal infrared data has been shown to provide more realistic estimates of thermophysical parameter uncertainties \citep{Durech17}.

\begin{figure*}
\includegraphics[width=2.0\columnwidth]{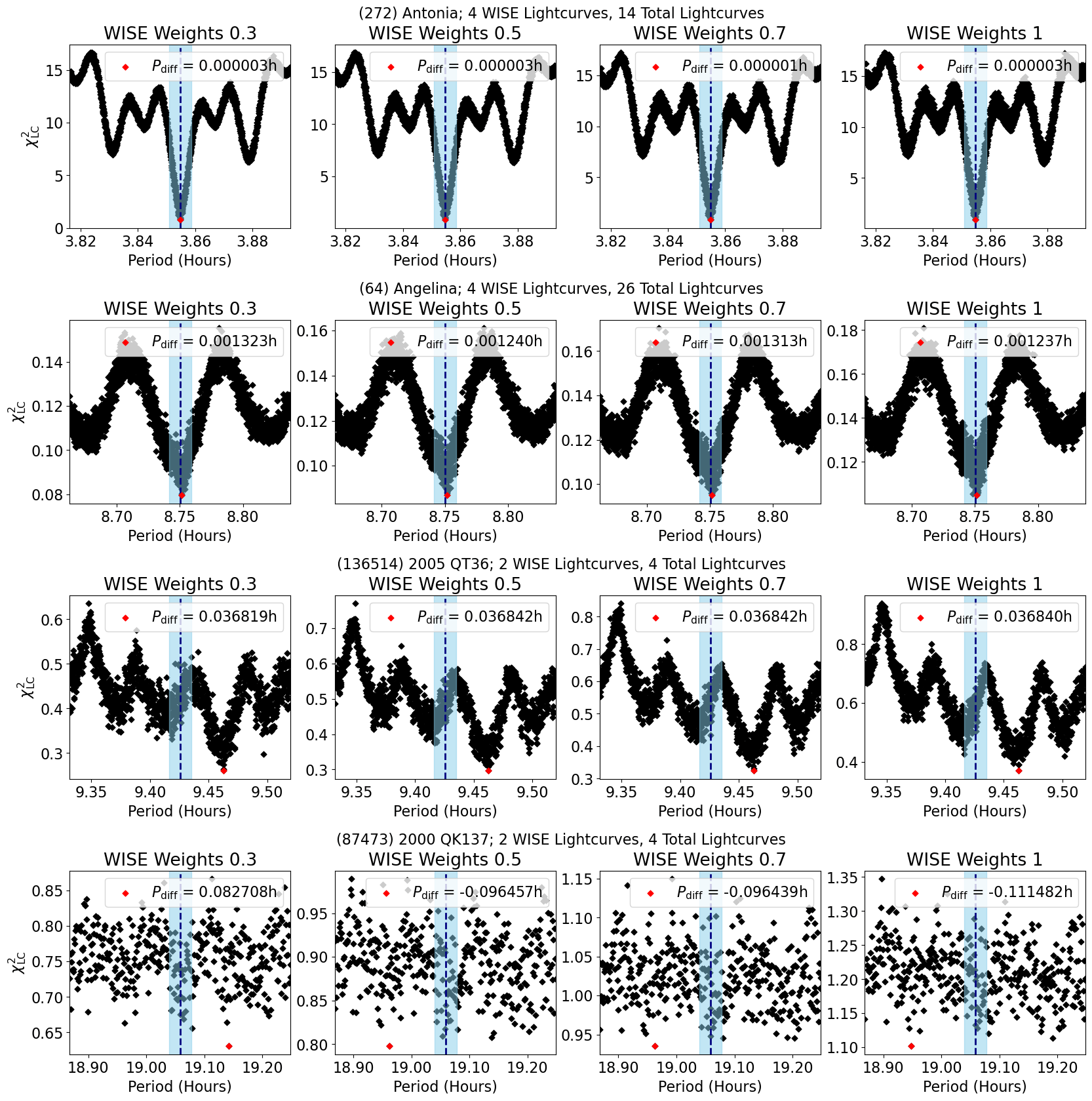} 
\caption{Example outputs of period searches conducted for (272) Antonia, (64) Angelina, (136514) 2005\,QT36, and (87473) 2000\,QK137. Each period search was conducted four times in total, using weights of 0.3, 0.5, 0.7, and 1 for the WISE lightcurves we supplemented to the DAMIT data set. The original DAMIT period is denoted by the vertical dashed line. The shaded region covers $\pm$0.1\% of the original DAMIT period. The period associated with the smallest $\chi_{\mathrm{LC}}^{2}$ in the search range is marked in red. The difference between it and the DAMIT period is labeled $P_{\mathrm{diff}}$ in each legend. The period search for (271) Antonia shows the ideal periodogram: a strong, tight V-shape with a clear single $\chi_{\mathrm{LC}}^{2}$ minimum. Its unique period was automatically identified as there were no other solutions with a $\chi_{\mathrm{LC}}^{2}$ under the threshold of 1.05 $\chi_{\mathrm{LC, min}}^{2}$. (64) Angelina is an example of a case where we manually identified the unique period. Like for (272) Antonia, the period search shows a clear V-shape and $\chi_{\mathrm{LC}}^{2}$ minimum. However, there were a few other solutions under the $\chi_{\mathrm{LC}}^{2}$ threshold with periods differing by a total range of 0.004 hour. (136514) 2005\,QT36 is an example of where a unique period was automatically identified, but we rejected the period for its large offset relative to the original DAMIT period---in this case by over 0.3\%. Finally, (87473) 2000\,QK137 is an example of a periodogram that is mostly flat and shows large scatter, and thus no unique period could be recovered.}
\label{fig.period_search}
\end{figure*}

We set the $\chi_{\mathrm{LC}}^{2}$ threshold in the period search at 5\% better than any other $\chi_{\mathrm{LC}}^{2}$ in the period search interval, while we use a 10\% cutoff for the finer pole search. If a unique period is automatically identified with our $\chi_{\mathrm{LC}}^{2}$ threshold, we visually inspect the periodograms and reject the period if it is inconsistent with the DAMIT solution (i.e., a period difference of greater than 0.1\%). For the period searches where no unique period is automatically identified, we take the period associated with the smallest $\chi_{\mathrm{LC}}^{2}$ if the periodogram has a clear single  minimum in $\chi_{\mathrm{LC}}^{2}$ and the period is consistent with the DAMIT solution. This manual inspection allows us to recover unique periods that happened to have a few other solutions that fell under our set $\chi_{\mathrm{LC}}^{2}$ threshold (Fig. \ref{fig.period_search}). Our imposed consistency check with the DAMIT solution is deliberately conservative. While the rotation periods in DAMIT may not necessarily be correct, we lack the compelling evidence in the form of new dense lightcurve data to support significantly discrepant results, and thus we err on the side of caution in our approach.

Though we use the same lightcurves as DAMIT to generate our shape models, DAMIT itself does not list any explicit weighting information. We thus adopt a simple scheme of treating the sparse survey telescope lightcurves with a weight of 0.1, while denser, more classical lightcurves are given a weight of 1. We give the WISE lightcurves a weight of 0.5, as their uncertainties, particularly in the $W1$ and $W2$ bands, are much larger than the uncertainties of dense optical lightcurves. We also run period searches using weights of 0.3, 0.7, and 1 with the WISE lightcurves to check for consistency, though the results are often largely identical outside of minor overall scaling differences in $\chi_{\mathrm{LC}}^{2}$. The WISE lightcurves have around an order of magnitude fewer data points than the dense optical lightcurves, so even if the weights were to be overestimated, the data do not dominate in the $\chi_{\mathrm{LC}}^{2}$ calculations. Though ideally the nonsurvey lightcurves should have hundreds of measurements over the entire rotational phase of the asteroid, we note that the existing lightcurves in DAMIT can vary wildly in terms of accuracy, period coverage, and sampling.

\section{Results}

\begin{figure*}
\centering
\includegraphics[width=2.0\columnwidth]{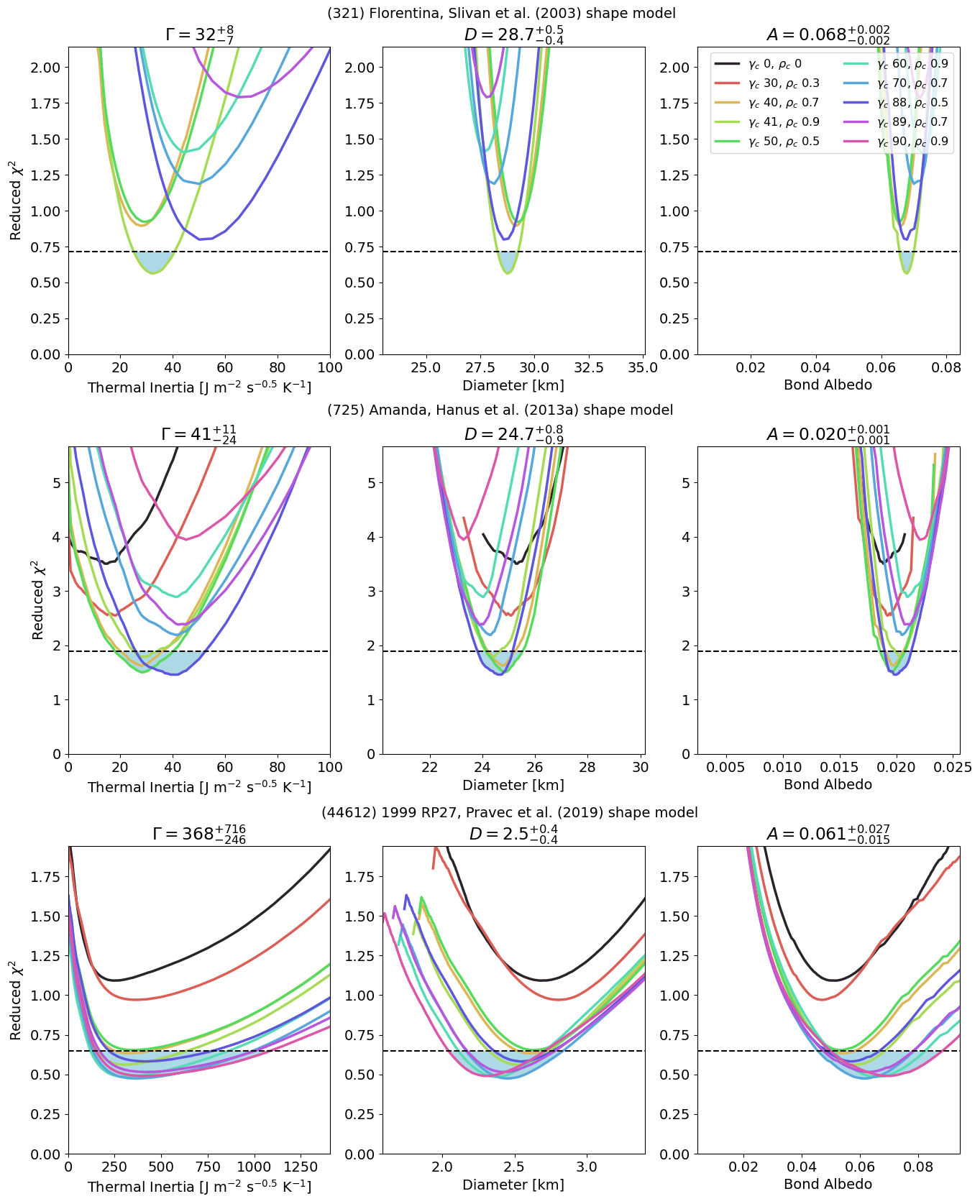} 
\caption{TPM example outputs for (321) Florentina, (725) Amanda, and (44612) 1999 RP27 and their respective best-fit thermal inertia, diameters, and Bond albedos found in Table \ref{tab.final}. Each curve represents a different roughness profile as a pair of the crater opening angle $\gamma_c$ and crater density $\rho_c$ as denoted in Table \ref{tab.roughness}. The dashed horizontal line is drawn at $\chi_{\mathrm{min}}^{2}$ (1 + $\sigma$), where $\sigma$ is related to the degrees of freedom $\nu$ by $\sigma$ = $\sqrt{2 \nu}/\nu$, and its intersection with the $\chi^2$ curves approximates the TPM fit's 1$\sigma$ uncertainty bounds of the best-fit thermal parameters found. The shaded region shows the contiguous regions around the $\chi^2$ minimum we use to identify the uncertainty ranges for each parameter.}
\label{fig.tpm_run}
\end{figure*}

\subsection{TPM Fits with DAMIT Shape Models}\label{sec.results.tpm}

\begin{figure}
\includegraphics[width=\columnwidth]{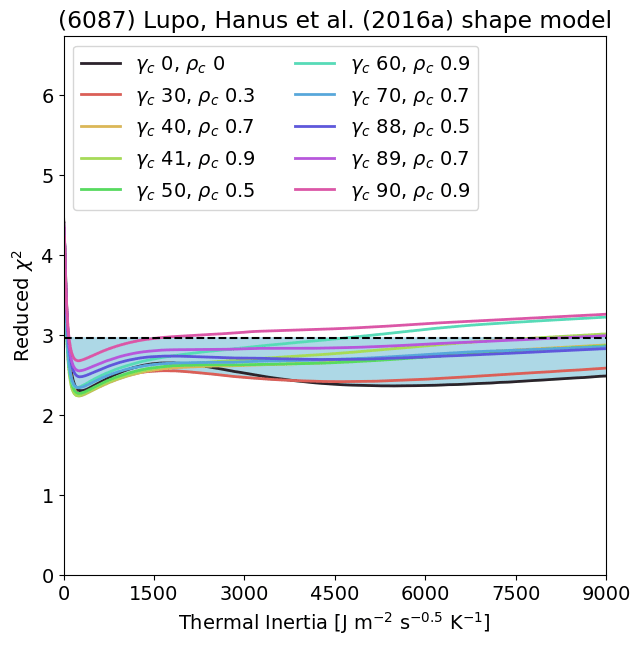} 
\caption{An example of what we consider a poorly constrained TPM fit for (6087) Lupo. Though the minimum reduced $\chi^2$ is not very large at 2.3, the overall flatness of the $\chi^2$ curve leaves the thermal inertia completely unconstrained.}
\label{fig.badfit}
\end{figure}

In the fully cryogenic WISE data, we identified a total of 97,657 asteroids with sufficient observations meeting our criteria in \S\ref{sec.tpminputs} for our thermophysical modeling, but only 2551 among them had existing DAMIT shape models. Of these, 1843 asteroids had two shape models, leaving us with a total of 4935 shape models to use in the TPM fitting. The vast majority of this set is made up of main-belt asteroids (MBAs), with a small number of Mars crossers, Trojans, and NEAs. For each asteroid, we record the minimum reduced $\chi^2$ of the grid of parameters tested in the TPM run (Fig. \ref{fig.tpm_run}). We run each TPM fit separately and independently in the cases of asteroids with multiple shape models. We can consider the thermophysical properties derived by the TPM to be well constrained if the TPM fit shows a clear minimum in its $\chi^2$ curve, even if the reduced $\chi^2$ is conventionally considered large (i.e., $>5$). Conversely, the thermophysical properties are unconstrained for TPM fits without a clear minimum, even if the reduced $\chi^2$ is small (Fig. \ref{fig.badfit}). Such cases are usually the result of poor thermal data (either too few measurements or large uncertainties), asteroid rotation periods greater than 15 hours, or close to pole-on observing geometries \citep{Hanus18}. A nonzero thermal inertia will introduce an offset between the points of maximum thermal absorption and emission, but this difference is small for slow rotators. In such cases, it is difficult to detect the thermal inertia signal, and therefore the offset is also difficult to constrain. Pole-on observations only show the same region on the asteroid, so the thermal lightcurve remains flat for any combination of thermophysical parameters, and so once again the parameters are difficult to constrain.

For our analysis, we reject all TPM fits where any thermal parameter was found to have zero uncertainty. The TPM code calculates zero and nonzero thermal inertia differently, and very rarely this presents itself as a conspicuous discontinuity in $\chi^2$ between thermal inertia of 0 and 1 J m$^{-2}$ s$^{-0.5}$ K$^{-1}$. Out of the 4935 shape models we applied the TPM to, a total of 12 fits were rejected in this manner. In each case, the shape model was derived from very sparse lightcurve data, which was often wholly comprised of only photometry from survey telescopes. In terms of $\chi^2$ cuts, we reject all TPM fits with reduced $\chi^2 > 10$. Additionally, we reject TPM fits with an overall flat distribution in $\chi^2$. For such cases, the thermal parameters are only poorly constrained at best. We identify these poorly constrained TPM fits by taking note of the ratio between the maximum and minimum $\chi^2$ in the TPM fit and rejecting fits where the ratio is less than 20. We measure this ratio based on the single roughness profile curve that gives the minimum $\chi^2$ in the thermal inertia range $\Gamma < 9000$ J m$^{-2}$ s$^{-0.5}$ K$^{-1}$ to remove the initial few steps of the TPM before it has reached a realistic albedo and size estimate. Of the 2551 asteroids we modeled with existing DAMIT shape models, 2152 met our initial reduced $\chi^2$ cutoff, while 1811 passed our $\chi^2$ ratio cut.

\begin{figure*}
\centering
\includegraphics[width=2.0\columnwidth]{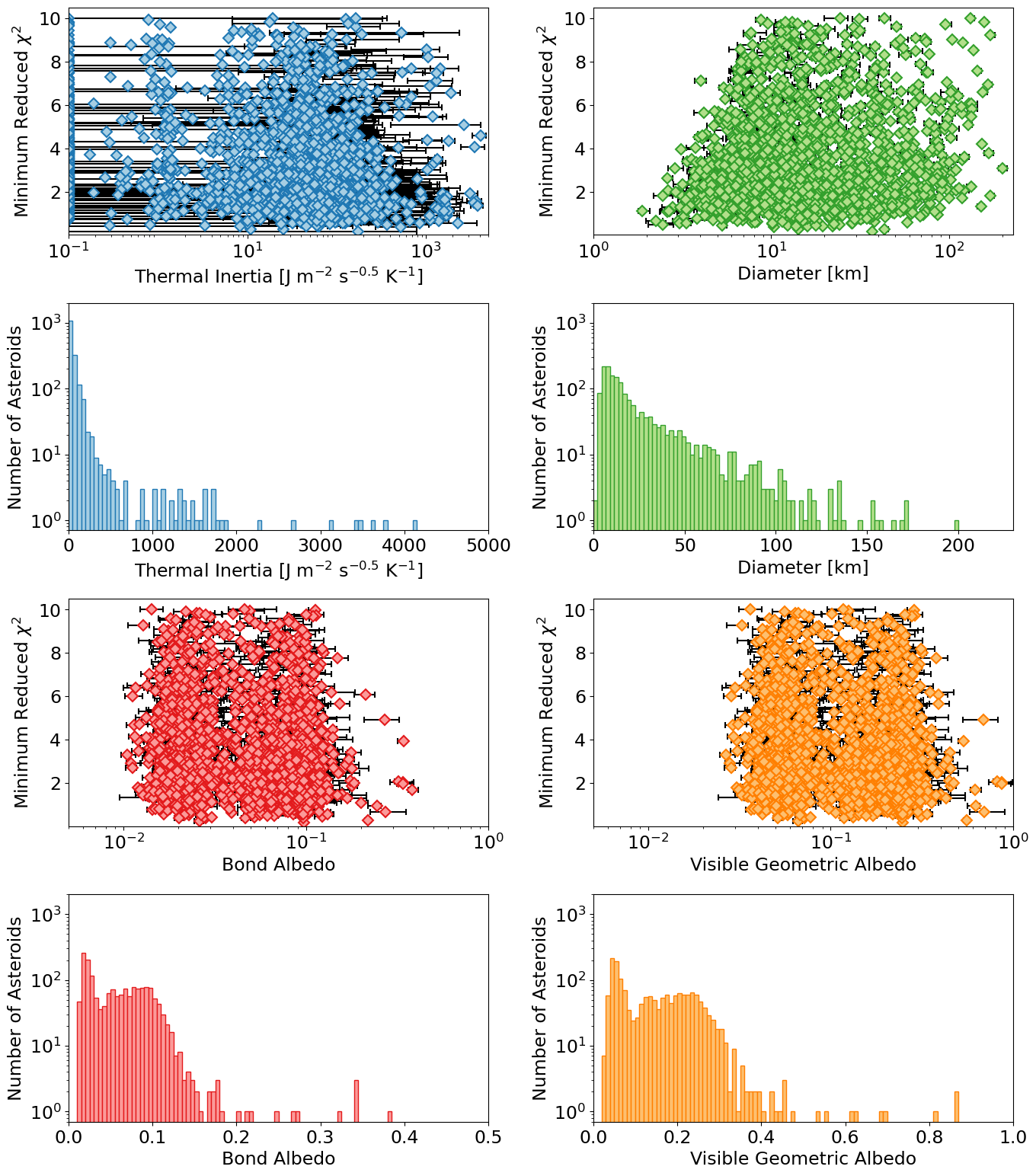} 
\caption{The thermal inertia, diameters, and albedos of our selected 1722 asteroids are shown with their uncertainties plotted against the reduced $\chi^2$ of their TPM fits. The distributions of each parameter, ignoring their uncertainties, are shown in the accompanying histograms. Asteroids with a best-fit thermal inertia of 0 J m$^{-2}$ s$^{-0.5}$ K$^{-1}$ are instead represented here with a value of 0.1 for the sake of the logarithmic scale. Error bars are only plotted for a best-fit thermal inertia of 50 J m$^{-2}$ s$^{-0.5}$ K$^{-1}$ or above in order to avoid visual clutter. We see in the top right panel that there appears to be a $\chi^2$ boundary that decreases with diameter, but this is primarily a consequence of the larger relative flux uncertainties in observations of smaller asteroids. The albedo distribution is clearly bimodal, with a sharp peak at around $A=0.03/p_V=0.06$ and a second peak at $A=0.09/p_V=0.23$. While there are a few thermal inertia values recorded over 1000 J m$^{-2}$ s$^{-0.5}$ K$^{-1}$, such cases should be treated with caution, as they are physically unrealistic and exceed nearly all other literature thermal inertia estimates. }
\label{fig.ensembles}
\end{figure*}

In Figure \ref{fig.ensembles}, we plot the fitted thermal inertia, diameters, and albedos for the total of 1722 asteroids in our sample that passed both of our $\chi^2$ cuts. In the cases where an asteroid has multiple shape models, we plot the derived parameters of the TPM fit associated with the smaller $\chi^2$. The visible geometric albedos are calculated from the Bond albedos using Equation \ref{eq.albedo}. The distributions of diameter and albedo are similar in shape to the distributions found in a much larger set of over 100,000 WISE-observed asteroids through NEATM modeling by \citet{Masiero11}, where the number of asteroids decreases with increasing diameter and the albedo distribution is bimodal. Though we have a fair number of TPM fits with $\Gamma > 1000$ J m$^{-2}$ s$^{-0.5}$ K$^{-1}$, such values are both physically unrealistic and much higher than most thermal inertia estimates in the literature, as we will discuss further in detail in \S\ref{sec.gamma_discuss}. We see no clear trend between each parameter and $\chi^2$ other than in diameter, where there appears to be a $\chi^2$ boundary that decreases with diameter. However, this is primarily a consequence of the larger relative flux uncertainties in WISE observations of smaller asteroids, which drops the $\chi^2$ in the TPM calculations.

\subsection{TPM Fits with Revised Shape Models}
\begin{figure*}
\centering
\includegraphics[width=2.0\columnwidth]{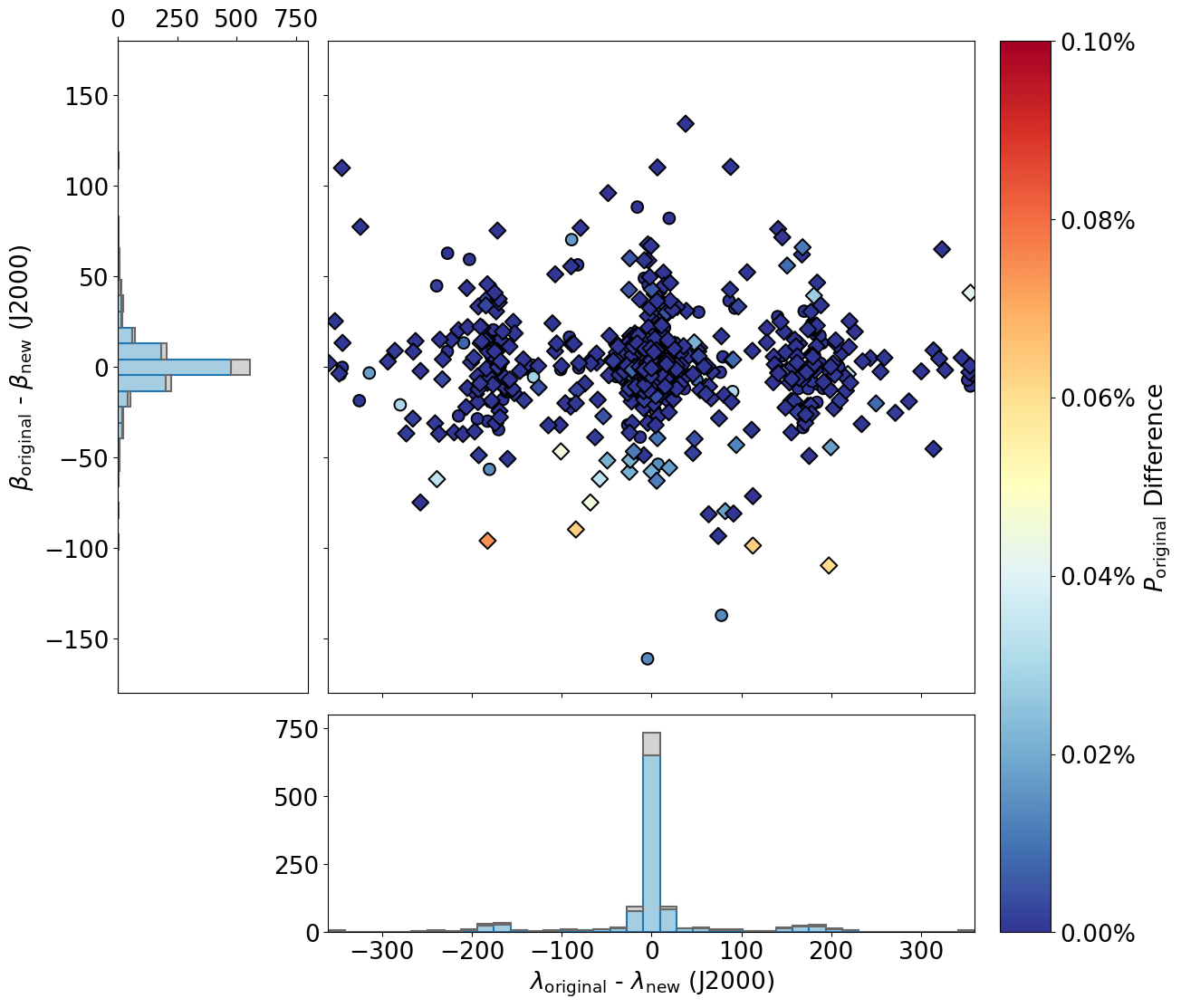} 
\caption{The differences in pole orientation in ecliptic coordinates and period between the original DAMIT shape models and revised shape models for each asteroid. The periods are reported as a percentage difference from the original DAMIT period. The diamond points and colored histograms represent revised shape models where the $\chi^2$ of the associated TPM fits fell into acceptable bounds by our criteria, while the circular points and grayscale histograms represent rejected TPM fits. We obtained a total of 1282 models for 696 asteroids. Of these, 63 asteroids had 1 DAMIT shape model and 1 revised shape model, 148 asteroids had 1 DAMIT shape model and 2 revised shape models, 43 asteroids had 2 DAMIT shape models and 1 revised shape model, and 440 asteroids had 2 DAMIT shape models and 2 revised shape models. In the cases where we had multiple DAMIT shape models, we only report the difference in the revised shape model with the closer DAMIT pole. The majority of revised shape models have poles which fall very closely to the DAMIT solutions with period differences of less than 0.01\%. Due to the ambiguities in the pole orientation, we also see the mirror solutions that show up in the local maxima in $\lambda$ at around $\pm180\degree$. }
\label{fig.spins}
\end{figure*}

In our initial TPM results, many of the asteroids have very high minimum $\chi^2$ values, suggesting that the TPM was unable to constrain any parameters with the given input data. One method available to us for improving the TPM results is to improve the shape models. We attempted to derive revised shape models for the 2551 asteroids in our sample by supplementing their lightcurves with WISE data following the methods outlined in \S\ref{sec.shape}. We obtained a total of 1282 new shape models for 696 asteroids. Due to the sparse lightcurve data available for the majority of our asteroids, we were often unable to resolve the ambiguities in an asteroid's pole orientation. In such cases, we have two derived shape models for a single asteroid, one each for each pole solution. Due to our selection criteria, the periods of our new shape models are very similar to their original DAMIT counterparts, though there are sometimes large differences in the pole orientation (Fig. \ref{fig.spins}).

\begin{figure}
\includegraphics[width=\columnwidth]{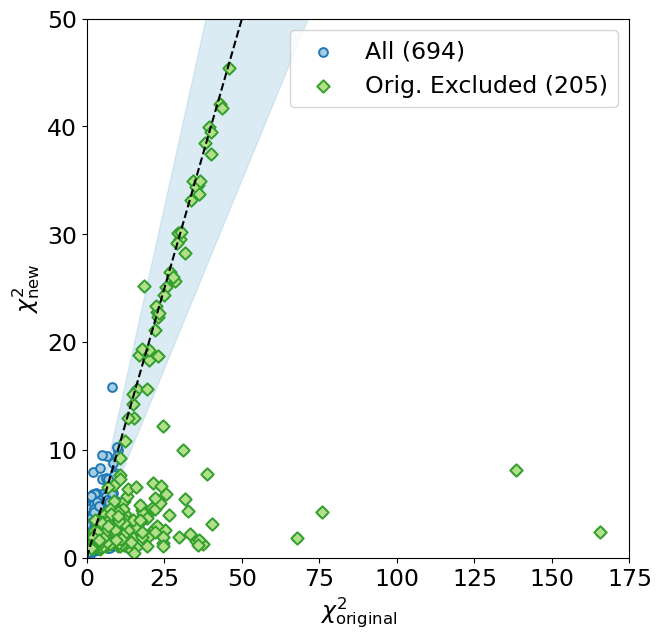} 
\caption{Our TPM fits' minimum reduced $\chi^2$ values, using both the nominal DAMIT shape models, $\chi_{original}^2$, and our new revised shape models, $\chi_{new}^2$. The points in green show asteroids that were initially removed with the $\chi^2$ cuts in \S\ref{sec.results.tpm}. The dashed line denotes where the two axes are equal. The shaded region shows where the values of the two axes are within 30\% of each other. Of the 694 asteroids with revised shape models, we saw improvements in the $\chi^2$ of their TPM fits for 540 asteroids compared to the fits using their DAMIT shape models. For the majority of asteroids, we either saw little change or slightly larger $\chi^2$ when switching to the revised shape models, but we also had several instances of dramatically improved TPM fits with the revised shape models. }
\label{fig.shape_chi2s}
\end{figure}

As with the DAMIT shape models, we again reject all TPM fits where any one thermal parameter was found to have zero uncertainty due to concerns with the TPM's boundary conditions, removing a total of five TPM fits from consideration, including the single revised shape model available for two asteroids, bringing our total number down to 694 unique asteroids. In Figure \ref{fig.shape_chi2s}, we plot our asteroids' $\chi^2$ of the TPM fits obtained with their revised shape models (where we use the best $\chi^2$ among them if there are multiple models for one asteroid) against the $\chi^2$ obtained from fits using their original shape models. The majority of asteroids appear to fall into either of two groups: one where there is little change between the two shape models and one where the improved reduced $\chi^2$ drops below 10, regardless of the original value. Even among the cases where the $\chi^2$ of the TPM fit worsened with the change in shape model, the $\chi^2$ was still less than 10 for the majority of asteroids. Of the 694 total asteroids, we saw improvements in $\chi^2$ in the TPM fit for 540 asteroids with their revised shape models. Among the asteroids with revised shape models, 205 had their TPM fits originally rejected by our $\chi^2$ cuts in \S\ref{sec.results.tpm}. The improvements in $\chi^2$ were enough to add 124 of these asteroids to our final sample of derived thermal parameters.

Our input data and results are given in an abridged form in Tables \ref{tab.final} and \ref{tab.inputs}, and the full data sets are available online on the VizieR database of astronomical catalogues at the Centre de Donn\'ees astronomiques de Strasbourg (CDS) website\footnote{\url{https://cds.u-strasbg.fr/}}. We report the TPM fit for each shape model used that passed our $\chi^2$ cuts. 

It is important to remember that the accuracy of TPM-derived physical properties will depend on the quality of the thermal flux measurements and shape models, particularly in the case of the thermal inertia \citep{Delbo15}. We advise readers to regard any atypical derived thermal inertia values among our results with caution. For example, we find the best-fit thermal inertia of several asteroids to be equal or close to zero. Zero thermal inertia, however, is unphysical in nature. Such derived values may be the product of inaccuracies in the TPM fitting, whether they be in the thermal flux measurements, the shape model, the $H$ and $G$ parameters, or the TPM itself. The derived thermal inertia value is also dependent on the assumed degree of surface roughness. While we note the best fitting \citet{Hapke84} mean surface slope for each TPM fit, we remind the reader that our roughness model is both very crude and not evenly sampled in parameter space, and thus should not be taken at face value. Moreover, the majority of asteroids in our sample were only observed by WISE in a single epoch, and the WISE observations for MBAs in general only cover a small total range in solar phase angles \citep[$14\degree$--$32\degree$;][]{Masiero11}. The thermal inertia and roughness may therefore share some degeneracy in the thermophysical modeling that we are unable to break due to the lack of infrared observations at multiple phase angles. Wider coverage in terms of wavelength, phase, rotational, and aspect angle would offer greater constraints for the TPM fitting that we do not have access to.

\begin{figure*}
\centering
\includegraphics[width=2.0\columnwidth]{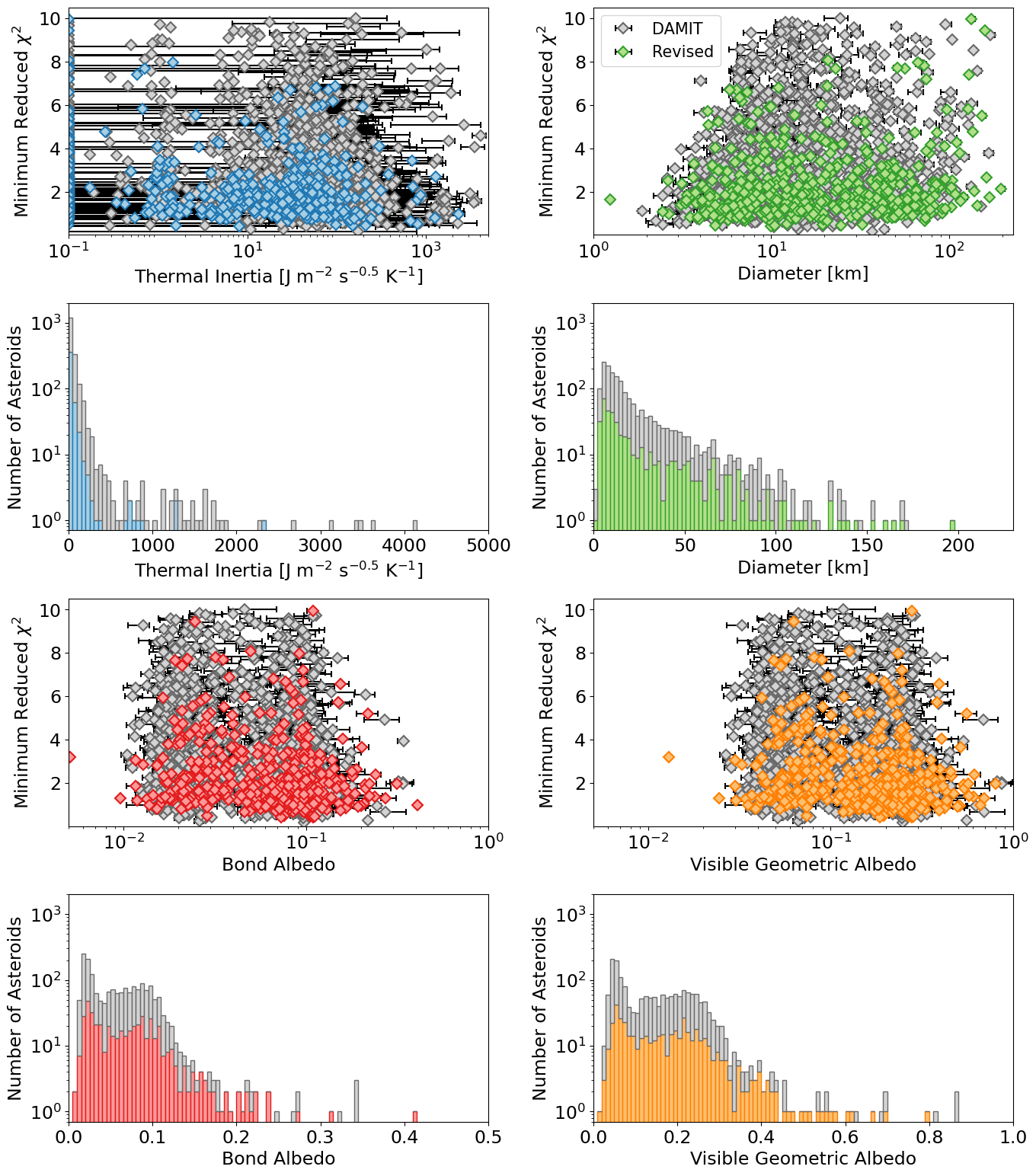} 
\caption{Thermal inertia, diameters, and albedos of our selected 1847 asteroid sample with their associated uncertainties. The visible geometric albedos are calculated from the Bond albedos using Equation \ref{eq.albedo}. As with Figure \ref{fig.ensembles}, asteroids with a best-fit thermal inertia of 0 J m$^{-2}$ s$^{-0.5}$ K$^{-1}$ are instead represented here with a value of 0.1 to accommodate the logarithmic scale. Error bars are only plotted for best-fit thermal inertia of 50 J m$^{-2}$ s$^{-0.5}$ K$^{-1}$ or above in order to avoid visual clutter. The DAMIT points in grayscale refer to the TPM fits using the original shape model and are unchanged from Figure \ref{fig.ensembles}. The colored points indicate the 476 cases in this sample where we obtained better TPM fits with the new shape models. Similarly, the colored portions of the histograms represent the fraction of total TPM fits that used the revised shape models. While the majority of the revised shape model TPM fits find thermal inertia values of less than 1000 J m$^{-2}$ s$^{-0.5}$ K$^{-1}$, the diameter and albedo values are more evenly distributed throughout their respective total ranges. The smallest asteroid shown in this figure is the 1.2 km NEA (1865) Cerberus, which is conspicuous here only as a result of the very few NEAs in our sample. The very low albedo at $A=0.005$ and $p_V=0.013$ belongs to the B-type MBA (1484) Postrema. Such low values for asteroid albedo are extremely rare \citep[see, e.g.,][]{Masiero11}, but our derived estimate agrees within 1$\sigma$ of the existing estimate derived through NEATM fitting of WISE data by \citet{Masiero14}.}
\label{fig.ensembles_best}
\end{figure*}

In Figure \ref{fig.ensembles_best}, we identify the best fit for each asteroid between the DAMIT and revised shape models by the $\chi^2$ of the fits, switching to using the revised shape model for 543 asteroids in our sample. Applying the same $\chi^2$ criteria we used in \S\ref{sec.results.tpm}, we now have 2239 asteroids with reduced $\chi^2 < 10$, 1914 asteroids with a $\chi^2$ ratio greater than 20, and 1847 asteroids meeting both conditions. Of these, we have 476 asteroids where the TPM fits used their revised shape models. The distributions of the thermal inertia, diameters, and albedos are largely unchanged from our original sample. The majority of the revised shape model TPM fits have a thermal inertia of less than 1000 J m$^{-2}$ s$^{-0.5}$ K$^{-1}$, which are values more in line with the literature thermal inertia estimates reported in Table \ref{tab.prev_gamma}. In contrast, the revised shape model TPM fits are more evenly distributed over the total ranges of the diameter and albedo.

\section{Discussion}\label{sec.discuss}
\begin{figure}
\includegraphics[width=\columnwidth]{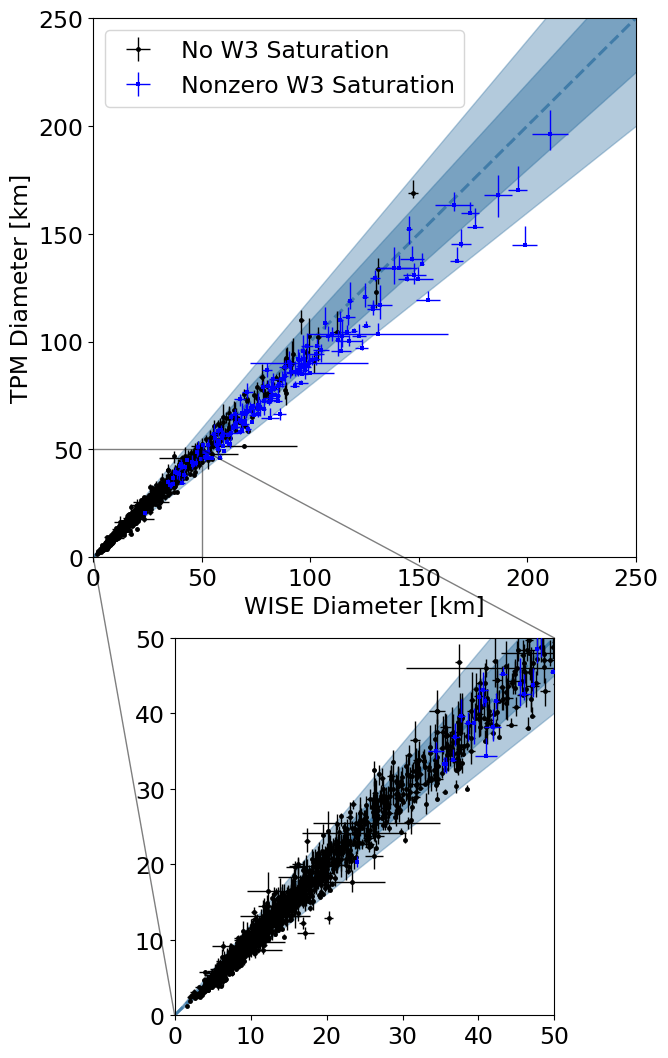} 
\caption{WISE diameters determined by NEATM and the diameters derived by our TPM fits. The inset shows an enlarged range between 0 and 50 km. The dashed line denotes where the two axes are equal, and the darker and lighter shaded regions show where the axes are respectively within 10\% and 20\% of each other. The asteroids plotted are separated based on the median saturation in their $W3$ band thermal flux measurements. We find that the 1652 nonsaturated asteroids have TPM diameters on average 95\% the size of the WISE diameters, while the difference for the 170 saturated asteroids is at 91\%. While our TPM fits only model the $W3$ and $W4$ bands, the NEATM fits have the benefit of also modeling the $W2$ band observation at high S/N, which downweights the significance of the $W3$ observation. We thus recommend readers to treat our derived thermophysical parameters for our $W3$ saturated asteroids with caution. }
\label{fig.diameters}
\end{figure}

\subsection{Comparison with NEATM Results}\label{sec.size}
Because many of the asteroids in our sample have previously estimated diameters, we can check the reliability of our results by comparing the diameters found by our best-fit TPM runs with the WISE diameters obtained with NEATM \citep{Harris98} and reported in the Planetary Data System \citep{Mainzer19}. Both sets of diameters were derived using the WISE thermal infrared data, so any significant discrepancies between the two arise from the differences in the models used, i.e., TPM versus NEATM. Of our 1847 asteroid sample, there are 1822 WISE diameters with which we can compare. We plot the WISE and TPM diameters against each other in Figure \ref{fig.diameters} where we see that 1746 asteroids fall within a 20\% agreement, and 1360 of those fall within 10\%. We expect that disagreements between the two mainly come about due to the nonrotating spherical shape model assumed by NEATM, which becomes a cruder approximation for more elongated asteroids.

However, we also see that the sizes of the larger asteroids, particularly above 100 km in diameter, show a slight, yet persistent underestimation by the TPM fits compared with the WISE diameters. Due to their size, these asteroids have the brightest thermal flux measurements, which causes partial saturation in the $W3$ band, leading to decreased accuracy in the reported magnitudes. The uncertainty on $W3$ for bright objects is known not to be dominated by random noise but by the systematic component from point-spread function (PSF)-wing fitting of saturated sources, which results in fluxes that are slightly overestimated \citep{Cutri12}. Among the 170 asteroids in our sample with a nonzero median $W3$ band saturation, the typical saturation is around 15\%, with the highest reaching just under 30\%. The $W3$ saturated asteroids have TPM diameters 9\% smaller than the WISE diameters on average, while the TPM diameters are only smaller by 5\% for the nonsaturated asteroids. Whereas in our TPM fits we only model the $W3$ and $W4$ bands, the NEATM fits for large asteroids typically also model the $W2$ band, which can shift the determined sizes \citep{Masiero14}. Because we do not make any correction for the partial $W3$ saturation nor do we use the measurements in the $W2$ band, we thus advise readers to treat our derived thermophysical parameters with caution for the $W3$ saturated asteroids in our sample.

\begin{figure}
\includegraphics[width=\columnwidth]{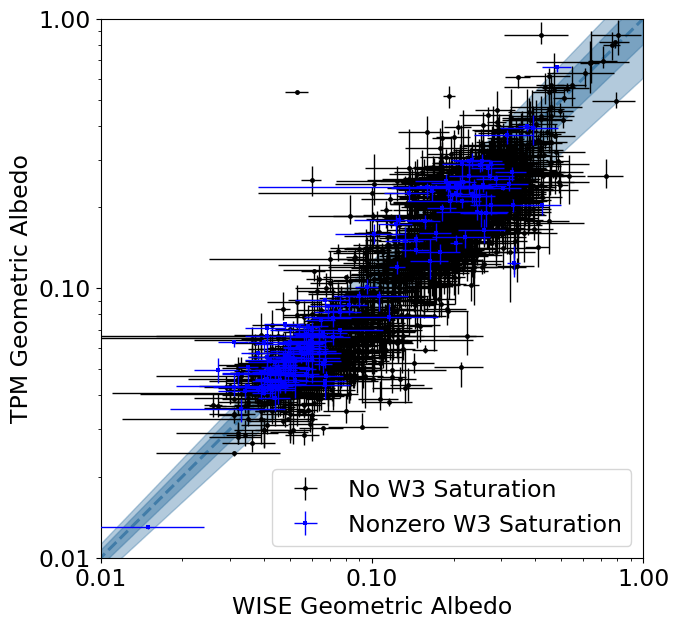} 
\caption{WISE albedos determined by NEATM and the albedos derived by our TPM fits, converted from Bond to visible geometric with Equation \ref{eq.albedo}. The dashed line denotes where the two axes are equal, and the darker and lighter shaded regions show where the axes are, respectively, within 20\% and 40\% of each other. There is a greater discrepancy between the two data sets relative to the diameter comparison in Figure \ref{fig.diameters}. The larger uncertainty on albedo is expected as it is proportional to $D^2$ and is also subject to the unreported uncertainties on the $H$ and $G$ parameters. We find that the 1652 asteroids without any saturation in their $W3$ band thermal flux measurements have albedos on average 98\% the value of the WISE albedos. The 170 saturated asteroids however are much higher, at 111\% the value of the WISE albedos. }
\label{fig.albedos}
\end{figure}

We can do the same comparison with the WISE NEATM-derived visible geometric albedos \citep{Mainzer19}, where we once again have 1824 values to compare with (Fig. \ref{fig.albedos}). We see markedly fewer matches between the two albedo estimates relative to the diameter comparison, with only 967 asteroids falling within a 20\% agreement and 496 within a 10\% agreement. The poorer agreement here is unsurprising due to the larger uncertainty on albedo. The albedo has about twice the uncertainty of the diameter as it is proportional to $D^2$ (Eq. \ref{eq.d}), and we will also have added contributions to the individual uncertainties from the unreported uncertainties on the $H$ and $G$ parameters. As with the diameters, the WISE albedos are also subject to the NEATM spherical shape model assumption. Whereas our $W3$ saturated asteroids were found to be 9\% smaller in size, we found albedos greater than the WISE values by about the same magnitude, at 11\%. For the nonsaturated asteroids, the albedos are smaller by only 2\% on average compared with the WISE albedos.

\begin{figure}
\includegraphics[width=\columnwidth]{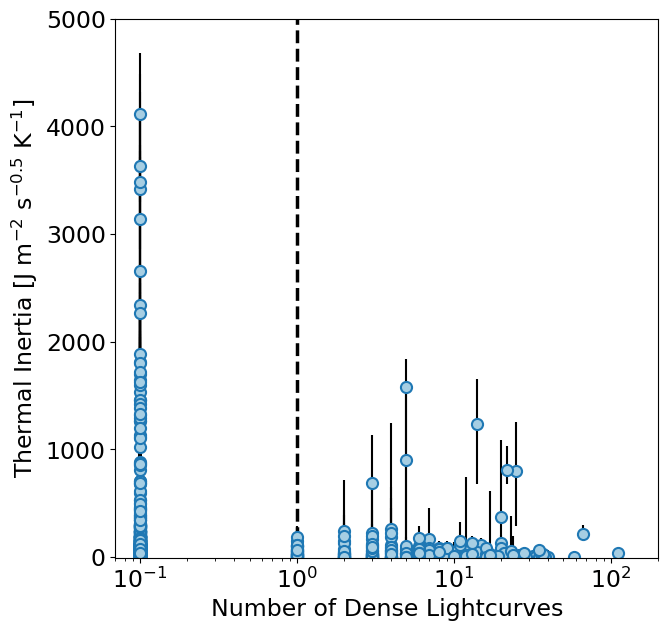} 
\caption{We plot the number of dense lightcurves included in each asteroid's optical lightcurve data used to derive their shape models against their thermal inertia estimates for our 1847 asteroid sample. Zero dense optical lightcurves are represented here as 0.1 to accommodate the logarithmic axis. The physically implausibly high thermal inertia estimates of $\Gamma > 1000$ J m$^{-2}$ s$^{-0.5}$ K$^{-1}$ are among the asteroids with the fewest optical lightcurve data. Limiting our sample to the shape models derived from data sets with at least one dense optical lightcurve, denoted by the vertical dashed line, is enough to remove nearly all cases with a best-fit thermal inertia of $\Gamma > 1000$ J m$^{-2}$ s$^{-0.5}$ K$^{-1}$.}
\label{fig.gamma_lcs}
\end{figure}

\subsection{Interpretation of Thermal Inertia}\label{sec.gamma_discuss}

\subsubsection{Relationship with Size}
While diameter and albedo can be fairly robustly determined in thermophysical modeling even with only an assumed spherical shape model, the same is not true for the determination of the thermal inertia. Shape models based on rich optical lightcurve data sets are closer to reality and tend to be smoother and have fewer planar regions. Shape models based only on sparse optical lightcurve data are crude approximations, often characterized by large planar facets that come together at sharp edges. Such shape models are physically unrealistic and thus are more likely to introduce inaccuracies into the thermophysical model. Additionally, asteroids in our sample with the highest thermal inertia estimates ($\Gamma > 1000$ J m$^{-2}$ s$^{-0.5}$ K$^{-1}$) are among those with the fewest optical lightcurve data. We consider these thermal inertia extremely high to the point of physical implausibility, and other thermophysical modeling studies have found very few equally high thermal inertia estimates (see Table \ref{tab.prev_gamma}). In order to ensure greater reliability from our thermal inertia analysis in this section, we limit our sample to the asteroids with shape models derived from data sets with at least one dense optical lightcurve, so chosen as to eliminate all of the physically implausibly high thermal inertia values (Fig. \ref{fig.gamma_lcs}). Here we define an optical lightcurve as dense if it satisfies three conditions: (1) includes at least 30 data points, (2) data points span 20\% or more of the asteroid's rotation phase, and (3) the mean separation between the data points is at most 2\% of the rotation phase. We consider lightcurves that meet all three of these conditions to be well sampled and free of large gaps in the portions of the rotation phase covered, and thus such lightcurves are more likely to yield more accurate shape models derived through lightcurve inversion. This cut reduces the sample down to TPM fits of 270 asteroids in total. We also exclude the TPM fits using partially saturated $W3$ observations, which further removes another 79 asteroids.

\begin{figure*}
\centering
\includegraphics[width=2\columnwidth]{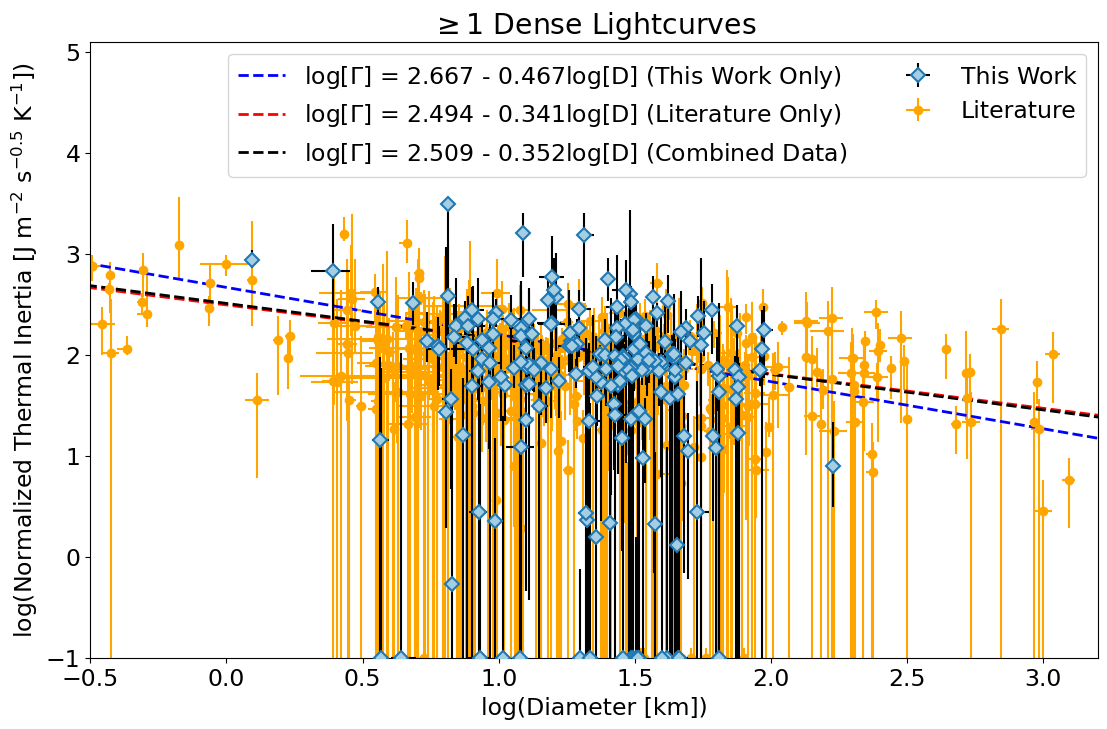} 
\caption{Our 191 asteroid sample's diameters and thermal inertia normalized to 1 au are shown here along with the 502 thermophysically derived literature values referenced in Table \ref{tab.prev_gamma}. The literature numbers include multiple instances of the same asteroid in order to account for the differences between independently derived parameters. Thermal inertia of 0 J m$^{-2}$ s$^{-0.5}$ K$^{-1}$ are represented here with values of 0.1 for the convenience of the logarithmic scale. When we apply linear fits to the data of the form $\log [\Gamma] = \alpha + \beta \log [D]$, we find the most similarity between the literature-only and combined data sets as we have nearly twice as many literature data points as we do of our sample. We also see that the fit to our sample alone prefers a larger $\alpha$ and steeper $\beta$ mainly due to the lack of large diameter data points, where we only have one larger than 100 km. Overall, we find good agreement between our results and the literature, with all terms within 1--2$\sigma$ of each other.}
\label{fig.diam_gamma}
\end{figure*}

In Figure \ref{fig.diam_gamma}, we plot the diameter against the normalized thermal inertia of our sample. As a function of the temperature $T$, thermal inertia will vary with heliocentric distance $r_{\mathrm{hel}}$ as $\Gamma \propto T^{3/2} \propto r_{\mathrm{hel}}^{-3/4}$ \citep{Delbo15}. To account for this temperature dependency in our comparisons, we normalize all thermal inertia values to $r_{\mathrm{h}} = 1$ au, using the median $r_{\mathrm{hel}}$ from all thermal observations of each asteroid. Although this relation assumes that all observations of an asteroid were conducted at similar $r_{\mathrm{hel}}$, of the 2551 asteroids in total we thermophysically modeled, 286 were observed by WISE in two epochs, and therefore at two different $r_{\mathrm{hel}}$. However, as the vast majority of our asteroids are MBAs, the differences in $r_{\mathrm{hel}}$ are typically small, so this only introduces a small inaccuracy. The largest change in $r_{\mathrm{hel}}$ in our sample is 0.7 au, and the total ranges for 70\% of the two epoch asteroids are smaller than 0.2 au.

The inverse correlation between size and thermal inertia was first identified by \citet{Delbo07}. With our sample limited to the 191 asteroids with dense optical lightcurve data and no saturation in the $W3$ observations, we confirm the dependence between diameter and thermal inertia to be broadly consistent with literature results (Table \ref{tab.prev_gamma}). Nearly 25\% of the 502 literature estimates listed are taken from \citet{Hanus18} who similarly modeled roughly 300 MBAs with cryogenic observations by WISE using the same thermophysical modeling code as in our work, albeit using slightly different data cuts. Our 191 asteroid sample includes 93 asteroids with a previously estimated thermal inertia in the literature, and 70 among them were estimated by \citet{Hanus18}, for which we largely see similar best-fit values and uncertainty ranges. 

We fit the data with a linear relation of the form $\log [\Gamma] = \alpha + \beta \log [D]$ using the curve\textunderscore fit routine in the Python library SciPy. $\Gamma$ values of zero are approximated as equal to 0.1 for the sake of the logarithmic convention. The uncertainties on the diameters are ignored as negligibly small for the fit. We run the fitting procedure several times to account for the asymmetric uncertainties on $\Gamma$. For the initial fit, we use the uncertainty of each data point averaged between the upper and lower bounds, which gives us our first $\alpha$ and $\beta$ values. We then use the linear fit to determine whether to use the upper or lower uncertainty bound for each data point. In other words, we take the upper uncertainty bound if the data point is lower than what is predicted by the fit, and vice versa. We then apply the fit again using these revised uncertainties. We repeat the fitting procedure in this manner until we see no more variation in the fit's $\alpha$ and $\beta$ values, which takes a total of three loops.

For our sample alone, we find $\alpha = 2.667 \pm 0.059$ and a slope of $\beta = -0.467 \pm 0.044$, with an associated reduced $\chi_{\mathrm{fit}}^2$ of 0.916. The literature estimates alone give $\alpha = 2.494 \pm 0.018$ and a slightly less shallow slope of $\beta = -0.341 \pm 0.013$, with an associated reduced $\chi_{\mathrm{fit}}^2$ of 1.664. Combining our data with the literature, we see $\alpha = 2.509 \pm 0.017$ and an overall slope very similar to the literature-only subset of $\beta = -0.352 \pm 0.012$, and an overall reduced $\chi_{\mathrm{fit}}^2$ of 1.455 (Fig. \ref{fig.diam_gamma}). While we see an offset of about 2$\sigma$ in both $\alpha$ and $\beta$ with our sample and the literature-only data set, the combined data set agrees within 1$\sigma$ of the literature-only data set. As there are around twice as many literature data points as there are data points from our sample, the literature data dominate in the fit of the combined data set. Our larger $\alpha$ and steeper $\beta$ terms with our sample's data compared to the literature-only data set can be explained by the lack of large asteroids in our sample, as we only have one data point with a diameter larger than 100 km. Though many of the literature estimates we fit use several different thermophysical models and thermal data sources, such as observations from the Spitzer Space Telescope \citep[e.g.,][]{Lamy08, Marchis12} and the Herschel Space Observatory \citep[e.g.,][]{O'Rourke12, Fornasier13, Ali-Lagoa20}, we see an overall strong consistency with our $\Gamma$--$D$ relation compared to the literature as a whole. Because our error bars are so large, however, it is difficult to draw any more definitive conclusions about thermal inertia's relationship with size. 

\begin{figure*}
\centering
\includegraphics[width=2\columnwidth]{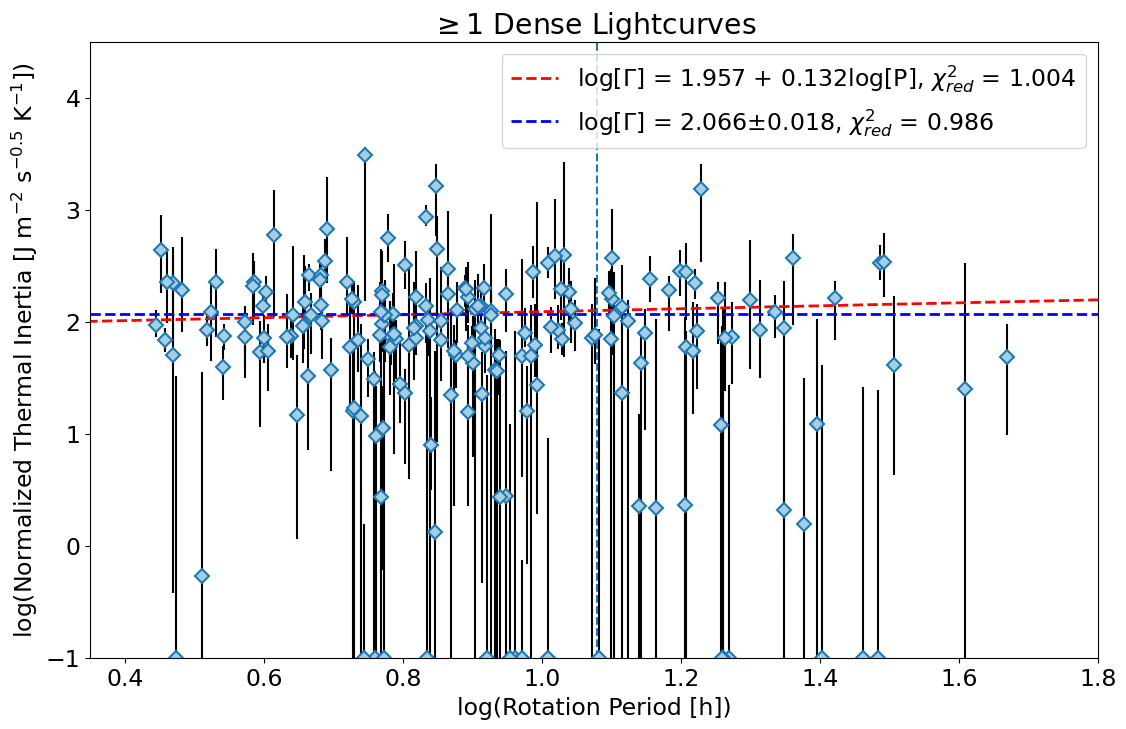} 
\caption{Our 191 asteroid sample's rotation periods are shown here plotted against their thermal inertia normalized to 1 au. Thermal inertia of 0 J m$^{-2}$ s$^{-0.5}$ K$^{-1}$ are represented here with values of 0.1 for the convenience of the logarithmic scale. The dashed vertical line shows the 12 hour period separation into the 143 fast and 48 slow rotators. By applying a linear fit and a constant fit to the data, we find that any evidence of a positive correlation is very weak and difficult to substantiate due to the small number of data points at $P>30$ h, the large scatter in the relation, and the large uncertainties on thermal inertia. }
\label{fig.per_gamma}
\end{figure*}

\subsubsection{Relationship with Rotation Period}
Some earlier studies \citep[e.g.,][]{Harris16} have suggested that slow rotators should present higher thermal inertia due to thermal observations being able to probe more deeply into an asteroid's surface. Other more recent studies \citep[e.g.,][]{Marciniak19, Ali-Lagoa20, Marciniak21} have found no excess of high thermal inertia values among slow rotators (conventionally defined as asteroids with $P>12$ hr), nor for low thermal inertia values among fast rotators. In our sample, we likewise find thermal inertia values greater than 100 J m$^{-2}$ s$^{-0.5}$ K$^{-1}$ to be generally present across our entire rotation period range, though we note that our sample is sparsely populated with rotation periods longer than 30 hours (Fig. \ref{fig.per_gamma}).

Applying a linear fit to the sample of the form $\log [\Gamma] = \alpha + \beta \log [P]$, we find $\alpha = 1.957 \pm 0.063$ and a small positive slope of $\beta = 0.132 \pm 0.074$, with an associated reduced $\chi_{\mathrm{fit}}^2$ of 1.004. However, the sample may in fact be better described by a constant. If we fix the slope to be zero, we instead find $\alpha = 2.066 \pm 0.018$, with an associated reduced $\chi_{\mathrm{fit}}^2$ of 0.986. Because of the small sample size of rotation periods longer than 30 hours, the large scatter in the data, and large uncertainties on the thermal inertia, however, we cannot reliably state whether there is a trend between thermal inertia and the rotation period or not.

\begin{figure}
\includegraphics[width=\columnwidth]{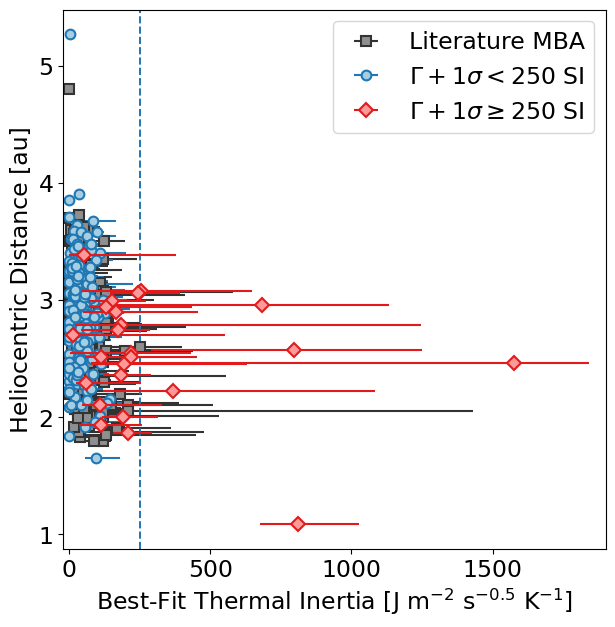} 
\caption{Best-fit thermal inertia for our 191 most reliable TPM fits with no $W3$ saturation listed in Table \ref{tab.final}, along with the 425 unique (or 442 total) MBAs reported in the literature (Table \ref{tab.prev_gamma}). Especially high thermal inertia cases are denoted by TPM fits where the upper uncertainty on the best-fit thermal inertia exceeds 250 J m$^{-2}$ s$^{-0.5}$ K$^{-1}$, marked by the vertical dashed line, which is equal to the \citet{Delbo09} thermal inertia estimate for (277) Elvira, the highest thermal inertia estimate reported for an MBA in Table \ref{tab.prev_gamma}. Our high thermal inertia estimates overlap in values with the estimates in the literature. However, our work can only provide tenuous evidence of unusually high thermal inertia in the main belt due to the consistently large associated uncertainties. }
\label{fig.big_gamma}
\end{figure}

\subsubsection{High Thermal Inertia in the Main Belt}

While MBAs have generally been found to have low thermal inertia (Table \ref{tab.prev_gamma}), there are over 30 literature thermal inertia estimates for MBAs higher than 100 J m$^{-2}$ s$^{-0.5}$ K$^{-1}$, albeit many with large uncertainties. The highest literature thermal inertia estimate reported for an MBA is $250\pm150$ J m$^{-2}$ s$^{-0.5}$ K$^{-1}$ for (277) Elvira \citep{Delbo09}. We have five MBAs with best-fit thermal inertia greater than 250 J m$^{-2}$ s$^{-0.5}$ K$^{-1}$ among our 191 most reliable TPM fits with no $W3$ saturation listed in Table \ref{tab.final}. If we include the upper uncertainty bound on the thermal inertia for this threshold, we have another 17 MBAs in the high thermal inertia range. In each instance, the thermal inertia is associated with a large uncertainty and is thus not very well constrained (Fig. \ref{fig.big_gamma}).

Indeed, for all but one MBA, the lower uncertainty bound on the thermal inertia is less than 250 J m$^{-2}$ s$^{-0.5}$ K$^{-1}$, placing the asteroids within the ranges of other MBA thermal inertia in the literature. The remaining MBA, (951) Gaspra, has a derived thermal inertia of $798^{+453}_{-508}$ J m$^{-2}$ s$^{-0.5}$ K$^{-1}$, which falls within the error bars of the estimate reported for (277) Elvira. Our other high thermal inertia case of interest is the Apollo NEA (1865) Cerberus, where we derive a best-fit thermal inertia of $809^{+219}_{-134}$ J m$^{-2}$ s$^{-0.5}$ K$^{-1}$, which is not atypical when compared with the thermal inertia of other NEAs in the literature.

While our derived high thermal inertia fits may not necessarily be discrepant with previous studies, it is possible that the high thermal inertia estimates for MBAs reported in the literature are themselves the results of poor TPM fits and untrustworthy. However, the high thermal inertia found in our sample are all associated with large uncertainties. Such values are thus best to be considered with caution. We do not recommend taking our findings as evidence of unusually high thermal inertia in the main belt.

\begin{figure*}
\centering
\includegraphics[width=2\columnwidth]{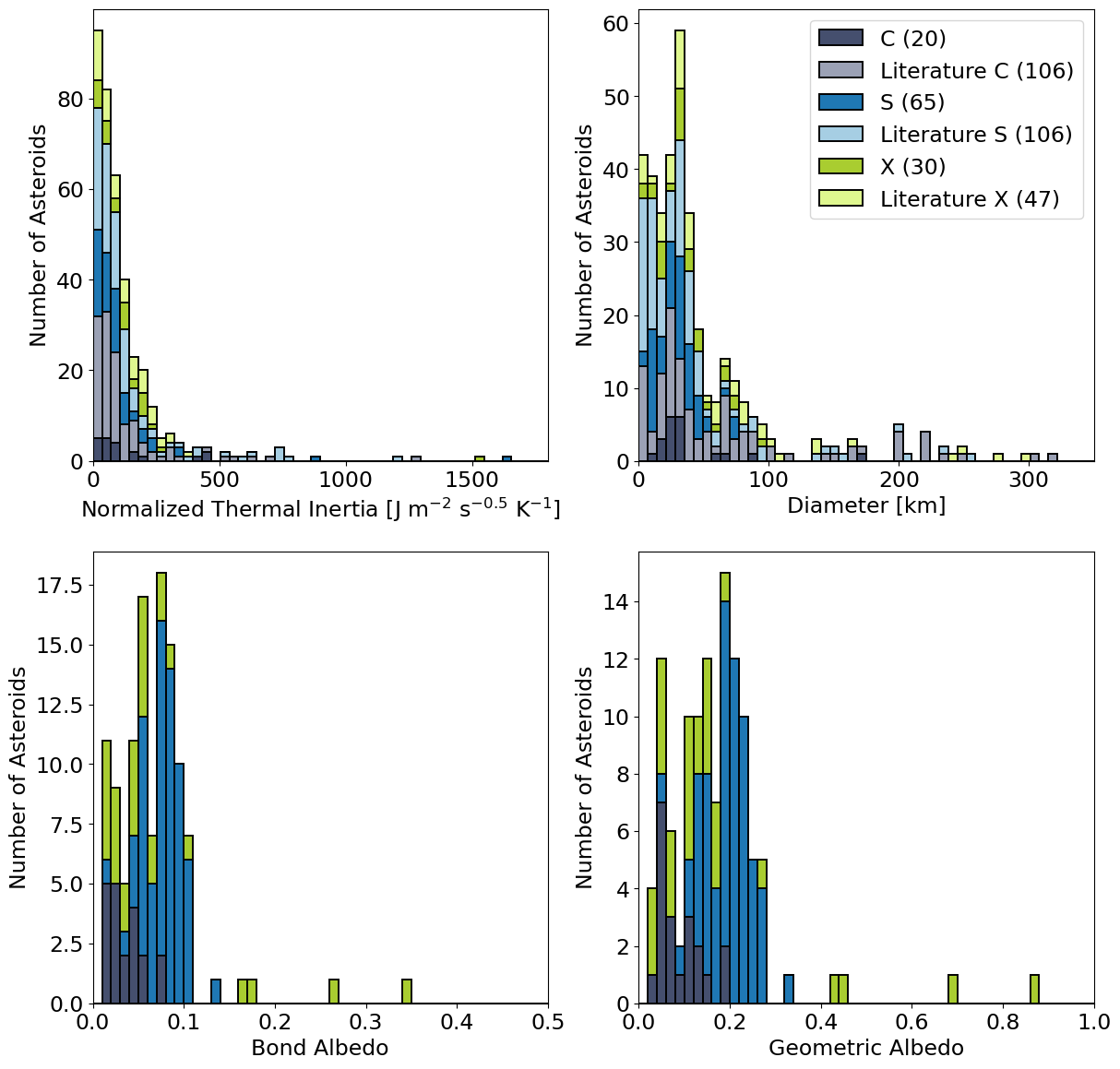} 
\caption{Thermal inertia (normalized to 1 au), diameter, and albedo represented as stacked histograms for our 191 asteroids with our most reliable TPM fits. The taxonomic types are sorted into three categories accordingly: C-, B-, F-, G-, D-, and T-types in the C-group, S-, V-, A-, Q-, R-, K-, and L-types in the S-group, and X-, M-, P-, and E-types in the X-group. The remaining 77 asteroids have unknown taxonomy and are not included here. Literature estimates for thermal inertia and diameter are included from Table \ref{tab.prev_gamma}. Note that the literature numbers count each duplicate asteroid in the table separately in order to account for the differences between independently derived parameters. The thermal inertia and size distributions are roughly equal when split by taxonomic class. The geometric albedo distribution is consistent with past observations of each taxonomic class, where C-group asteroids are most common among albedos of $p_V < 0.10$ and S-group asteroids are most common for brighter albedos, while X-group asteroids are scattered across the entire albedo range \citep[e.g.,][]{Gradie82, Tholen84}. }
\label{fig.taxplots}
\end{figure*}

\subsection{Thermal Parameters and Taxonomy}

To assess how the thermal parameters might differ depending on taxonomy, we sort the taxonomic classes into three main categories: C-group asteroids which include C-, B-, F-, G-, D-, and T-types, S-group asteroids which include S-, V-, A-, Q-, R-, K-, and L-types, and X-group asteroids which include X-, M-, P-, and E-types. For our 191 asteroids with the most reliable TPM fits (i.e., at least one dense optical lightcurve and no saturation in the thermal flux measurements), we have 20 C-group asteroids, 65 S-group asteroids, 30 X-group asteroids, and 77 asteroids with unknown taxonomy (Fig. \ref{fig.taxplots}). For these groupings, the Small Main-Belt Asteroid Spectroscopic Survey (SMASS) II \citep{Bus02} taxonomy is preferentially used over the \citet{Tholen84} taxonomy where available. Some asteroids in the \citet{Tholen84} taxonomy have multiple spectral types assigned due to ambiguities in the color analyses. In these cases, we adopt the best-fitting spectral type, which denoted by the first letter in the sequence. 

Taxonomy appears to be no strong predictor of thermal inertia or size, with each group of asteroids being about as equally common over the same parameter space for both parameters derived in our work and in the literature estimates. In principle, metal-rich regolith should present higher thermal inertia due to the higher thermal conductivity of the material, as can be seen in iron meteorites compared with ordinary and carbonaceous chondrites \citep{Opeil10}. However, taxonomy alone is far from a strong indicator of metal content. Although M-types are traditionally regarded as the metallic taxonomic class, the results of radar observations have suggested metal-poor compositions for some M-type asteroids \citep[e.g.,][]{Magri99, Magri07, Shepard15}, while spectroscopic studies of M-type asteroids have found cases of spectral features inconsistent with metallic compositions \citep[e.g.,][]{Clark04, Hardersen05}. Relatively high radar albedos, which are consistent with high metal content, have also been found for some S- and C-type asteroids, suggesting that metal-rich compositions may not be limited to M-type objects \citep[e.g.,][]{Magri07}. It is important to note, however, that the effects of composition are difficult to distinguish from other factors that will affect the thermal inertia, such as the average grain size or degree of compaction \citep{Gundlach13}. A high thermal inertia may not necessarily be the product of a metal-rich surface.

The geometric albedo distribution is consistent with the previous observations of each group of asteroids, where C-group asteroids are most common for albedos of $p_V < 0.10$, while S-group asteroids are most common for $p_V > 0.10$ \citep[e.g.,][]{Gradie82, Tholen84}. There are a few instances of unusual albedos per taxonomic class that may warrant further investigation into whether or not they were correctly classified.

We have two S-group asteroids with albedos fainter than $p_V = 0.10$ in our sample. The lowest albedo S-group asteroid is (1332) Marconia, which is classified as an Ld-type under the SMASS II \citep{Bus02} taxonomy. We estimate its geometric albedo to be $p_V = 0.044^{+0.002}_{-0.005}$, which is several sigma fainter than the WISE NEATM-derived estimate of $p_V = 0.063 \pm 0.008$, reported in the Planetary Data System \citep{Mainzer19}. The second lowest albedo S-group asteroid is (673) Edda, which is classified as an S-type under both the SMASS II \citep{Bus02} and \citet{Tholen84} taxonomies. We estimate its geometric albedo to be $p_V = 0.097^{+0.001}_{-0.002}$, which agrees within 1$\sigma$ of the WISE NEATM-derived estimate of $p_V = 0.092 \pm 0.034$. 

We have three C-group asteroids with albedos brighter than $p_V = 0.15$. (1317) Silvretta has an estimated geometric albedo of $p_V = 0.199^{+0.057}_{-0.024}$, which is slightly fainter than the WISE NEATM-derived estimate of $p_V = 0.275 \pm 0.047$. (1317) Silvretta may however be better described as an X-type. The classification for (1317) Silvretta is particularly tenuous. Its full \citet{Tholen84} type is given as CX:, where the trailing colon signifies an uncertain classification. Our other bright C-group albedos belong to (390) Alma and (531) Zerlina. (390) Alma is denoted as a DT-type in the \citet{Tholen84} taxonomy. Our geometric albedo estimate of $p_V = 0.188^{+0.015}_{-0.015}$ is several sigma fainter than the WISE NEATM-derived estimate of $p_V = 0.268 \pm 0.011$. (531) Zerlina is a B-type under the SMASS II \citep{Bus02} taxonomy. Unlike the other bright C-group albedos, our geometric albedo estimate of $p_V = 0.151^{+0.009}_{-0.008}$ for this asteroid is several sigma brighter than the WISE NEATM-derived estimate of $p_V = 0.101 \pm 0.007$.

X-group asteroids cover a wide range in reflectivity and can be split into three types in the \citet{Tholen84} taxonomy based on albedo: P ($p_V < 0.10$), M ($0.10 < p_V < 0.30$), and E ($p_V > 0.30$). These subdivisions in albedo are cleanly recovered for the cases where a particular one of these types was specified. Our four P-types had albedos between $0.03 < p_V < 0.05$, our six M-types had albedos between $0.11 < p_V < 0.27$, and our three E-types had albedos between $0.42 < p_V < 0.69$.

\subsection{Yarkovsky Follow-up}\label{sec.yark}

\begin{figure}
\includegraphics[width=\columnwidth]{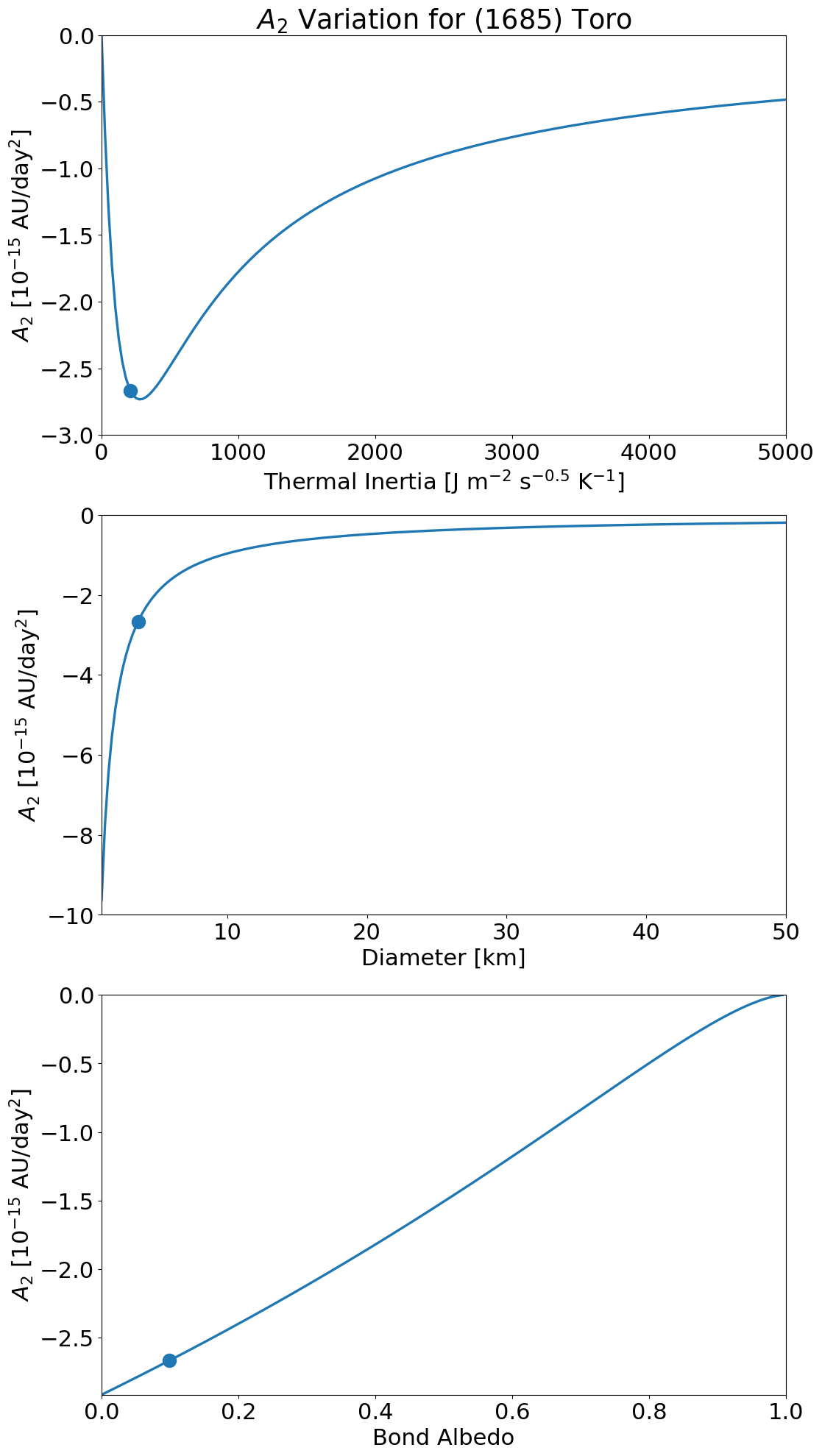} 
\caption{An example of how the transverse acceleration parameter $A_2$ changes as a function of thermal inertia, diameter, and Bond albedo respectively for (1685) Toro. Parameters not being varied are otherwise held constant at the best-fit values our TPM found for the asteroid, denoted by the single scatter point in each panel. While $A_2$ is inversely proportional with diameter and albedo, its magnitude drops off both for very low and very high thermal inertia values.}
\label{fig.yark_params}
\end{figure}

If an asteroid had no thermal inertia, it would absorb and release heat from the Sun immediately, making its temperature distribution symmetrical around its subsolar point. The net force acting on the asteroid would thus only be radially outward from the Sun. With a thermal inertia component, however, there is a delay between the thermal absorption and emission. This Yarkovsky force introduces a transverse component to the net force that will induce a secular change to the asteroid's semimajor axis. However, the magnitude of the Yarkovsky effect also decreases at very high thermal inertia. If no heat is transferred over one rotation cycle, the temperature distribution is once again uniform and the Yarkovsky effect drops to zero. Moreover, while having some thermal inertia is necessary for the Yarkovsky effect, its magnitude also depends on an asteroid's heliocentric distance, axial tilt, size, shape, and rotation period \citep{Bottke06}. The majority of asteroids in our sample have diameters under 30--40 km, which is the maximum size range where we can theoretically expect asteroids to undergo possible Yarkovsky acceleration \citep{Vokrouhlicky15b}. At larger sizes, the acceleration induced by the Yarkovsky effect becomes vanishingly small (Fig. \ref{fig.yark_params}). Because the Yarkovsky effect decreases with increasing heliocentric distance and larger asteroid size, the vast majority of asteroids with a measured nongravitational transverse acceleration parameter ($A_2$) are subkilometer NEAs.

As of this writing, there are published $A_2$ values for 244 NEAs on the JPL Small-Body Database. Yarkovsky drift rate detections for an additional 149 NEAs have also been reported in \citet{Greenberg20}. The largest reported diameters among these asteroids belong to the 6.25 km (3200) Phaethon and 5.4 km (4179) Toutatis \footnote{\url{https://ssd.jpl.nasa.gov/tools/sbdb_query.html}}. (3200) Phaethon's size was derived from 2007 and 2017 radar data from Arecibo and Goldstone \citep{Taylor19} but may be overestimated. Past thermophysical modeling efforts have derived a diameter of 5.1 km for (3200) Phaethon \citep{Hanus16}, and more recent work using the same radar data set in combination with stellar occultations and multiple apparition lightcurves has revised the size down to 5.3 km \citep{Marshall21}. In our sample, only (1685) Toro has an $A_2$ value explicitly determined\footnote{\url{https://ssd.jpl.nasa.gov/tools/sbdb_lookup.html\#/?sstr=1685}}, where $A_2 = -3.339 \pm 0.6433 \times 10^{-15}$ au day$^{-2}$. We also note that one other asteroid in our sample, (1865) Cerberus, has a measured orbit-averaged semimajor axis drift rate of $\left \langle da/dt \right \rangle = -3.75 \pm 1.8 \times 10^{-4}$ au My$^{-1}$ \citep{Greenberg20}, which is related to $A_2$ by Equation 5 in \citet{Farnocchia13}

\begin{equation}\label{eq.drift}
\left \langle da/dt \right \rangle = \frac{2 A_2 (1 - e^2)}{n} \left ( \frac{1\,\mathrm{au}}{r_{\mathrm{hel}}} \right )^d
\end{equation}

\noindent where $e$ is the eccentricity, $n$ is the mean motion, and $r_{\mathrm{hel}}$ is the heliocentric distance. $d$ is a value that depends upon the object's thermal properties. For the extreme cases, $d$ can be as low as 0.5 or as high as 3.5, though it is more typically between 2 to 3 \citep{Farnocchia13}.

In order to identify the most promising Yarkovsky candidates for observational follow-up, we compute a rudimentary Yarkovsky acceleration prediction for every asteroid in our sample using Equation 10 in \citet{Farnocchia13}

\begin{equation}\label{eq.yark}
A_2 = \frac{4(1-A)}{9} \Phi (1\,\mathrm{au}) f(\Theta) \cos (\gamma)
\end{equation}

\noindent where $A$ is the Bond albedo and $\gamma$ is the spin obliquity. $\Phi (1\,\mathrm{au})$ is the standard radiation force factor, equal to

\begin{equation}
\Phi (1\,\mathrm{au}) = \frac{3 G_S}{2 \rho D c}
\end{equation}

\noindent where $G_S$ is equal to 1361 W m$^{-2}$ \citep{Kopp11} and is the solar constant at 1 au, $\rho$ is the bulk density, $D$ is the mean diameter, and $c$ is the speed of light. $f(\Theta)$ is the function of the thermal parameter $\Theta$, which is equal to

\begin{equation}
f(\Theta) = \frac{0.5 \Theta}{1 + \Theta + 0.5 \Theta^2}
\end{equation}

\noindent $\Theta$ is given by

\begin{equation}
\Theta = \frac{\Gamma}{\epsilon \sigma_{SB} T_*^3} \sqrt{\frac{2 \pi}{P}}
\end{equation}

\noindent where $\Gamma$ is the thermal inertia, $\epsilon$ is the bolometric emissivity, $\sigma_{SB}$ is the Stefan-Boltzmann constant, and $P$ is the rotational period. $T_*$ is the subsolar temperature, given by

\begin{equation}
T_* = \sqrt[4]{\frac{(1-A)G_S}{\epsilon \sigma_{SB} p^2}}
\end{equation}

\noindent $p$ is the semi latus rectum, given by

\begin{equation}
p = a(1-e^2)
\end{equation}

\noindent where $a$ is the semimajor axis and $e$ is the eccentricity.

For each asteroid, we estimate the uncertainty ranges on $A_2$ by choosing values of $\Gamma$, $D$, and $A$ in their respective 1$\sigma$ uncertainty bounds such that they minimize or maximize $A_2$. As in \S\ref{sec.tpm}, we assume $\epsilon = 0.9$. We use the bulk densities reported in \citet{Carry12} if they are not denoted as unreliable, but such estimates are only available for fewer than 300 asteroids. For the asteroids in our sample that do not have existing estimates for $\rho$, if the SMASS II \citep{Bus02} taxonomy is known, we assume values equal to the average bulk densities with 20\% accuracy in Table 3 of \citet{Carry12}. If only the \citet{Tholen84} taxonomy is known, we use the average bulk densities of \citet{Krasinsky02} following their classification scheme. The average bulk densities for each scenario is given in Table \ref{tab.yarkdens}. In the cases where the taxonomic class is unknown, we reference the typical visible geometric albedo ranges for each composition type and treat the asteroid as a C-type for $p_V < 0.10$ and an S-type otherwise \citep{Tholen84}, then apply the \citet{Krasinsky02} average bulk densities.

\begin{deluxetable}{ccc}
\tablecolumns{3}
\tablecaption{Average Bulk Density by Taxonomic Class\label{tab.yarkdens}}
\tablehead{\colhead{Taxonomic Class} & \colhead{Bulk Density $\rho$} & \colhead{Number of Asteroids} \\
		   \colhead{} & \colhead{[g\,cm$^{-3}$]} & \colhead{}
		   }
\startdata
\cutinhead{SMASS II \citep{Bus02} Taxonomy}
S   & 2.72 & 105 \\
Sq  & 3.43 & 13 \\
B   & 2.38 & 18 \\
C   & 1.33 & 30 \\
Cb  & 1.25 & 4 \\
Ch  & 1.41 & 23 \\
X   & 1.85 & 35 \\
Xc  & 4.86 & 10 \\
Xe  & 2.60 & 10 \\
Xk  & 4.22 & 11 \\
K   & 3.54 & 12 \\
V   & 1.93 & 4 \\
\cutinhead{\citet{Tholen84} Taxonomy}
C, D, P, T, B, G, F & 1.38 & 46 \\
S, K, Q, V, R, A, E & 2.71 & 86 \\
M & 5.32 & 9 \\
\enddata
\tablecomments{Average bulk densities based on taxonomic class, adapted from Table 3 in \citet{Carry12} for the SMASS II \citep{Bus02} classes and Table 1 in \citet{Krasinsky02} for the \citet{Tholen84} classes. In the latter case, several classes were grouped together and regarded respectively as compositional C- and S-types. The \citet{Carry12} bulk densities were given in three forms for the level of accuracy considered: 20\%, 50\%, and no precision restriction; we adopt the 20\% accuracy values here. Note that not every SMASS II taxonomic class has a listed density. For each row, we also give the number of asteroids where we took the associated average bulk density. The \citet{Krasinsky02} densities are only used in cases where the \citet{Carry12} densities were not available. These numbers do not include asteroids where the taxonomic class was not available and only assumed based on the geometric albedo. }
\end{deluxetable}

\begin{figure*}
\centering
\includegraphics[width=1.6\columnwidth]{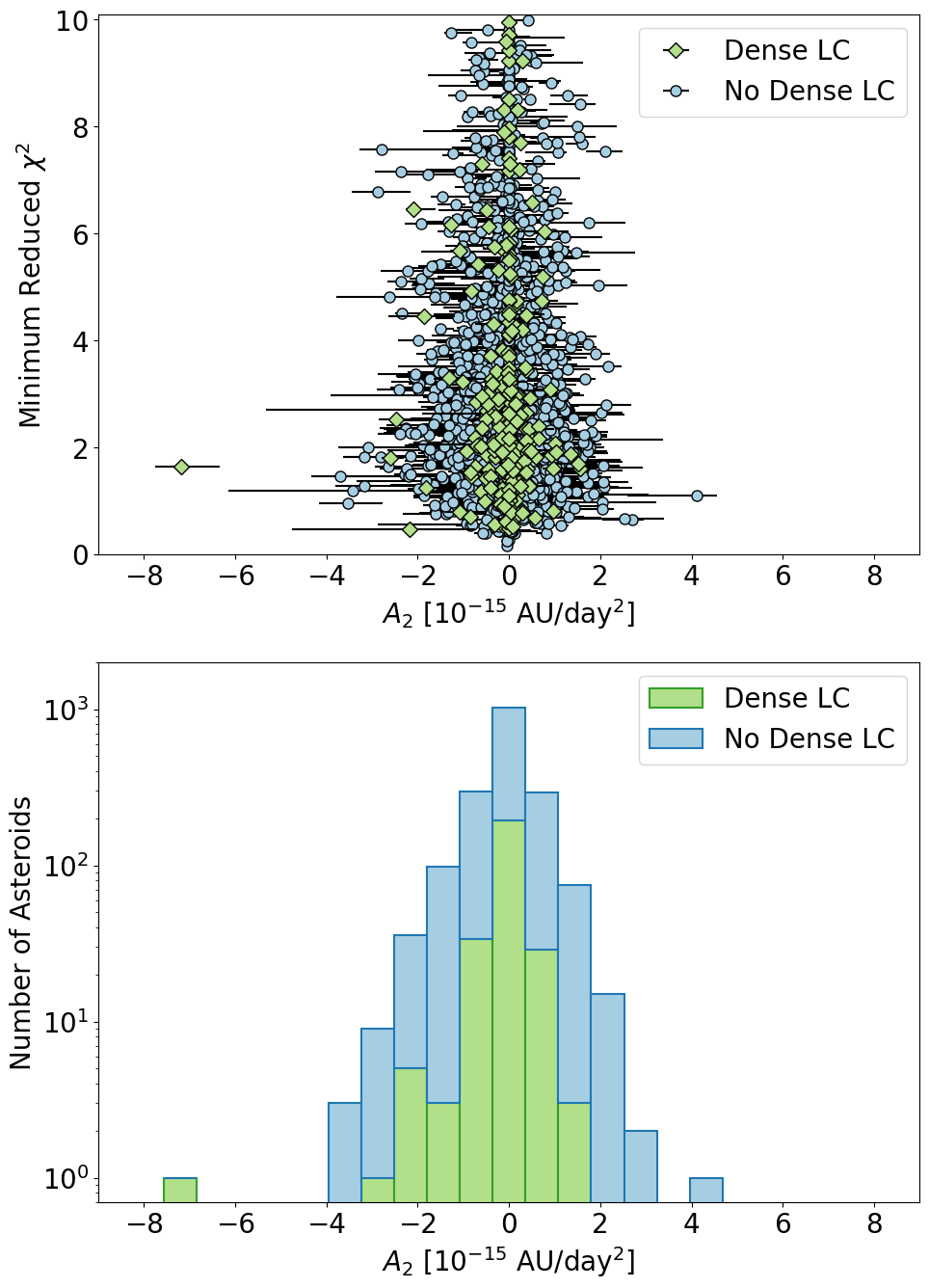} 
\caption{The top panel plots the estimated nongravitational transverse acceleration parameter $A_2$ for every asteroid in our sample against the $\chi^2$ of their best TPM fit, while the bottom panel shows the histogram of the same data. We estimated an $A_2$ value for each asteroid following the procedure outlined in \S\ref{sec.yark}. The data points are distinguished between asteroids which, for the purposes of deriving their shape models, had at least one dense optical lightcurve in their data set and those that did not. In our sample, we have 57 asteroids with an $A_2$ larger in magnitude than the smallest published $A_2$ value on the JPL Small-Body Database ($A_2 = 1.9 \times 10^{-15}$ au day$^{-2}$), with 10 also having at least 1 dense optical lightcurve. These asteroids are the most promising for direct observational follow-up to confirm their Yarkovsky drift rates. However, because the lower bound on thermal inertia is unconstrained for 11 of these 57 best Yarkovsky candidates, the lower bound on $A_2$ here is also unconstrained, and as such the drift rates of these asteroids may be undetectably small. }
\label{fig.yark}
\end{figure*}

With our simplified approach, we found an $A_2$ estimate of $-2.6^{+0.3}_{-0.2} \times 10^{-15}$ au day$^{-2}$ for (1685) Toro, which is within a 1$\sigma$ agreement with the JPL Small-Body Database value of $-3.339 \pm 0.6433 \times 10^{-15}$ au day$^{-2}$. We compile the 57 asteroids in our sample with an acceleration greater in magnitude than $A_2 = 1.9 \times 10^{-15}$ au day$^{-2}$ in Table \ref{tab.yark} using the best-fit values for $\Gamma$, $D$, and $A$. This cutoff in $A_2$ is the same as the smallest published $A_2$ value on the JPL Small-Body Database, which is for the asteroid (35107) 1991\,VH. Among the asteroids with dense optical lightcurve data, we have a total of four asteroids, 1 NEA and 3 MBAs, other than (1685) Toro in this parameter space. These asteroids are the best candidates in our sample for observational follow-up to measure their orbital drifts directly with astrometry. If successful, independent Yarkovsky measurements can also help further constrain the thermal inertia values derived through thermophysical modeling. However, we caution the reader that 11 of these 57 Yarkovsky candidate asteroids have an unconstrained lower bound on their thermal inertia, and so the lower bound on their $A_2$ estimate is similarly unconstrained. These asteroids may have undetectably small drift rates in reality (Fig. \ref{fig.yark}).

\section{Conclusions and Future Prospects}\label{sec.conclusion}

Using thermal observations from WISE \citep{Wright10, Mainzer11} taken during its fully cryogenic phase and convex shape models from DAMIT \citep{Durech10}, we have thermophysically modeled a total of 2551 asteroids. With the addition of WISE photometry to the optical lightcurve data sets for each asteroid, we were able to derive revised shape models for nearly 700 asteroids through lightcurve inversion, producing 540 TPM fits with an improved $\chi^2$ compared with the DAMIT shape models. We set two conditions for what fits to include in our final sample: a reduced $\chi^2 < 10$ and a $\chi^2$ ratio of at least 20. The purpose of the second condition is to quantify what we consider a well-defined minimum in the $\chi^2$ curve of the TPM fit which is needed to constrain the derived thermal parameters. Our final sample consists of TPM fits for 1847 asteroids.

Currently, there are published thermal inertia estimates for only a few hundred asteroids (Table \ref{tab.prev_gamma}). Thanks largely in part to the sharp increase in asteroid shape models in recent years, we have derived through thermophysical modeling thermal inertia estimates for 1847 asteroids, though only 191 have both sufficient optical lightcurve data and nonsaturated thermal flux measurements for the derived thermal inertia to be reliable. However, even within this limited sample, we have 98 asteroids that did not previously have a thermal inertia estimate in the literature, which increases the number of asteroids with thermal inertia estimates by around 20\%. 

We found our derived diameters and albedos to be broadly consistent with the NEATM-derived values with the WISE thermal data set \citep{Mainzer19}. Our derived thermal inertia were broadly consistent with the trends found with diameter \citep[e.g.,][]{Delbo07} and rotation period \citep[e.g.,][]{Marciniak19, Ali-Lagoa20, Marciniak21} in the literature, though in both cases the large uncertainties on thermal inertia limit the conclusions we were able to draw from the data. We applied a linear fit to the diameters and thermal inertia normalized to 1 au of the form $\log[\Gamma]=\alpha+\beta\log[D]$, finding best-fit values of $\alpha = 2.667 \pm 0.059$ and $\beta = -0.467 \pm 0.044$ for our sample alone and $\alpha = 2.509 \pm 0.017$ and $\beta = -0.352 \pm 0.012$ when combined with literature estimates. We applied a linear fit to the rotation periods and normalized thermal inertia and found a slightly positive trend between the two parameters. However, we also found that the data were similarly well described by a constant. Given the small sample size of rotation periods longer than 30 hours, the large scatter in the data, and large uncertainties on the thermal inertia, we were unable to reliably state whether there is a trend between thermal inertia and the rotation period or not. We found the thermal parameters when split by asteroid taxonomic class to be about equally distributed in terms of thermal inertia and diameter, and the albedo distribution was consistent with past studies \citep[e.g.,][]{Gradie82, Tholen84}. Finally, we used our derived thermal parameters to compute a rudimentary Yarkovsky acceleration for every asteroid in our sample, identifying the 57 best candidates for confirmation via observational follow-up.

Of the 2551 asteroids we thermophysically modeled, 71\% use shape models derived by \citet{Durech18b,Durech19, Durech20}, which take advantage of using incidental sparse photometry from survey telescopes such as the Asteroid Terrestrial-impact Last Alert System \citep[ATLAS;][]{Tonry18} instead of having to conduct time-intensive targeted campaigns to obtain dense optical lightcurves for individual asteroids. We can anticipate similar success with the large volume of data expected from upcoming all-sky surveys, such as the Vera Rubin Observatory Legacy Survey of Space and Time \citep[LSST;][]{lsst}. LSST is a ground-based survey with full science operations planned by 2023. Its field of view covers nearly 10 square degrees of sky. It will cover the entire visible sky twice a week by taking nearly a thousand 30 s observations each night. With a light-gathering power equivalent to that of a 6.7 m diameter primary mirror, its limiting magnitude will eclipse those of current all-sky surveys such as the 0.5 m telescopes used by ATLAS and the 1.8 m Panoramic Survey Telescope and Rapid Response System (Pan-STARRS1) telescope \citep{panstarrs}. The photometric data gathered by all-sky surveys will be bolstered by the European Space Agency mission Gaia\footnote{\url{https://www.cosmos.esa.int/web/gaia}} \citep{Gaia16}, which was launched in 2013 December with the goal of obtaining accurate astrometry and photometry for more than a billion stars. By scanning the entire sky, Gaia will also incidentally observe thousands of asteroids. DR2 \citep{Gaia18} contained astrometric and photometric data for around 14,000 asteroids. Although the number of measurements per individual asteroids was small, DR2 has already enabled derivations of hundreds of new shape models \citep{Durech19}. We can expect a greater number of more accurate shape models with the upcoming Gaia DR3 catalog, which is anticipated to become available on 2022 June 13.

Thermophysical modeling benefits greatly from the additional constraints granted by having multiple epochs of observations in order to probe the thermal emission across the asteroid's diurnal temperature distribution. The seven-month fully cryogenic WISE mission, however, was limited to observing asteroids at a maximum of two epochs, and of our 2551 asteroids, 89\% were only observed at one epoch. Thermal infrared data of asteroids have traditionally been limited in quantity due to the challenges of observing at such longer wavelengths from the ground. However, we will soon see much more plentiful thermal data with the upcoming Near-Earth Object Surveyor\footnote{\url{https://neos.arizona.edu/}}, a space-based survey with a planned launch in 2026. Its two infrared channels, which span 4--5.2 and 6--10 $\mu$m respectively, were designed to constrain the typical thermal emission curve expected for NEAs. The telescope will be passively cooled over its planned five-year mission, thus requiring no cryogen. Within its nominal mission length, the telescope is expected to detect millions of asteroids, with the majority having several epochs of observations.

In the interest of saving significant computational time and labor, we do not account for the uncertainties in the shape models, and thus the error bars on our derived parameters are essentially lower limits. The effects of incorporating shape uncertainty into thermophysical modeling have been largely unexplored. \citet{Hanus15} introduced a method of generating several shape models for each asteroid by bootstrapping the photometric data used for lightcurve inversion, which was later applied to a larger sample of roughly 300 WISE-observed asteroids by \citet{Hanus18}. Asteroids where the TPM solutions differed among the set of shape models pointed to the shape model as the primary source of uncertainty, which was the case for roughly 35\% of their asteroid sample. Improvements to the shape model can mainly come about with more data, such as additional optical photometry.

One of the longstanding challenges of thermophysical modeling is the lack of benchmark tests to verify the accuracy of the derived parameters. The earliest TPMs were made to model thermal observations of the lunar surface and were later adapted by \citet{Spencer89} for general planetary bodies with an assumed spherical shape. The most commonly used thermophysical models used for asteroids today \citep[e.g.,][]{Lagerros96, Delbo07, Rozitis11} are all variations of the \citet{Spencer89} model. Lunar regolith is made up of fine grains and has an extremely low thermal conductivity \citep{Winter71}, which is very different from the surfaces of many asteroids, particularly for the smallest in size. Ground truth information can be highly valuable as an independent check for TPM-derived parameters, but missions to individual asteroids are so far very few in number due to their enormous costs and several years of planning required. Meteorite studies are more accessible but do not investigate the same environment as TPMs, as asteroid fragments are unlikely to survive unchanged through ejection from their parent body and atmospheric entry. The orbital perturbation induced by the Yarkovsky effect is dependent on thermal inertia, and direct measurement of an asteroid's orbital drift can serve to place constraints on the TPM-derived thermal inertia. However, this is complicated by the Yarkovsky drift's dependency on several other physical parameters that we may not have any knowledge of, such as the asteroid's bulk density. In the near future, we will see more and more asteroid thermal data and shape models become available, which will translate to a growth in the number of asteroids with thermophysically derived parameters. Large-scale independent verification of such parameters, however, is likely a long way off yet.

\section*{Acknowledgements}

{\footnotesize We would like to thank the anonymous referees for the valuable and thorough comments. The work of D.H. and D.J.T. are supported by NASA grant Nos. NNX13AI64G and 80NSSC21K0807. The work of J.H. has been supported by INTER-EXCELLENCE grant LTAUSA18093 from the Czech Ministry of Education, Youth, and Sports and by the grant GA20-04431S of the Czech Science Foundation. This publication makes use of data products from the Wide-field Infrared Survey Explorer, which is a joint project of the University of California, Los Angeles, and the Jet Propulsion Laboratory/California Institute of Technology, funded by the National Aeronautics and Space Administration. This publication also makes use of data products from NEOWISE, which is a project of the Jet Propulsion Laboratory/California Institute of Technology, funded by the Planetary Science Division of the National Aeronautics and Space Administration. The technical support and advanced computing resources from University of Hawaii Information Technology Services -- Cyberinfrastructure, funded in part by the National Science Foundation MRI award \#1920304, are gratefully acknowledged. }

\bibliographystyle{aasjournal}
\bibliography{tpmwise}

\clearpage

\section{Appendix}

\startlongtable
\begin{deluxetable*}{rlcccccc}
\centering
\tablecaption{Literature TPM/ATPM-derived Properties\label{tab.prev_gamma}}
\tablehead{
	\multicolumn2c{Asteroid} & \colhead{$D$} & \colhead{$\Gamma$} & \colhead{Taxonomy} & \colhead{$r_{\mathrm{h}}$} & \colhead{Orbit} & \colhead{Reference} \\ 
	\multicolumn2c{} & \colhead{(km)} & \colhead{(J\,m$^{-2}$\,s$^{-0.5}$\,K$^{-1}$)} & \colhead{} & \colhead{(au)} & \colhead{} & \colhead{}
	}
\startdata
1 & Ceres & $923^{+20}_{-20}$ & $10^{+10}_{-10}$ & C & 2.767 & MBA & [1] \\ 
1 & Ceres & $951^{+9}_{-7}$ & $25^{+15}_{-10}$ & C & 2.80 & MBA & [2] \\ 
2 & Pallas & $544^{+43}_{-43}$ & $10^{+10}_{-10}$ & B & 2.772 & MBA & [1] \\ 
2 & Pallas & $536^{+5}_{-5}$ & $30^{+15}_{-15}$ & B & 2.95 & MBA & [2] \\ 
3 & Juno & $234^{+11}_{-11}$ & $5^{+5}_{-5}$ & S & 2.671 & MBA & [1] \\ 
3 & Juno & $252^{+2}_{-3}$ & $60^{+25}_{-20}$ & S & 2.70 & MBA & [2] \\ 
4 & Vesta & $525^{+1}_{-1}$ & $20^{+15}_{-15}$ & V & 2.3 & MBA & [3] \\ 
4 & Vesta & $520^{+12}_{-6}$ & $35^{+55}_{-23}$ & V & 2.36 & MBA & [2] \\
\enddata
\end{deluxetable*}

\clearpage
\startlongtable
\begin{deluxetable*}{rlcccccccc}
\centering
\tablecaption{Final Sample TPM-derived Properties\label{tab.final}}
\tablehead{
\twocolhead{Asteroid} & \colhead{$D_{\mathrm{WISE}}$} & \colhead{$D$} & \colhead{$\Gamma$} & \colhead{$A$} & \colhead{$p_V$} & \colhead{$\bar{\theta}$} & \colhead{$\chi^2_{\mathrm{red}}$} & \colhead{Model ID} \\ 
\colhead{} & \colhead{} & \colhead{(km)} & \colhead{(km)} & \colhead{(J\,m$^{-2}$\,s$^{-0.5}$\,K$^{-1}$)} & \colhead{} & \colhead{} & \colhead{$\degree$} & \colhead{} & \colhead{}
} 
\startdata
5 & Astraea* & 108.3$\pm$3.7 & $104.8^{+7.8}_{-6.2}$ & $45^{+58}_{-40}$ & $0.106^{+0.015}_{-0.015}$ & $0.271^{+0.038}_{-0.038}$ & 55.4 & 6.7 & D1 \\
 & & & $102.4^{+4.8}_{-2.1}$ & $10^{+19}_{-10}$ & $0.113^{+0.004}_{-0.010}$ & $0.288^{+0.010}_{-0.025}$ & 55.4 & 2.1 & R1 \\
 & & & $105.2^{+4.3}_{-2.9}$ & $14^{+20}_{-14}$ & $0.107^{+0.005}_{-0.008}$ & $0.272^{+0.013}_{-0.021}$ & 55.4 & 2.1 & R2 \\
14 & Irene* & 140.8$\pm$8.4 & $132.6^{+6.6}_{-0.0}$ & $0^{+29}_{-0}$ & $0.113^{+0.000}_{-0.010}$ & $0.287^{+0.000}_{-0.025}$ & 55.4 & 10.0 & D1 \\
 & & & $134.3^{+5.9}_{-0.0}$ & $0^{+21}_{-0}$ & $0.110^{+0.000}_{-0.009}$ & $0.279^{+0.000}_{-0.022}$ & 55.4 & 9.9 & R2 \\
17 & Thetis* & 84.9$\pm$2.0 & $73.1^{+1.0}_{-0.0}$ & $0^{+28}_{-0}$ & $0.096^{+0.000}_{-0.002}$ & $0.244^{+0.000}_{-0.006}$ & 55.4 & 1.2 & D1 \\
 & & & $72.6^{+1.3}_{-0.0}$ & $0^{+30}_{-0}$ & $0.096^{+0.000}_{-0.003}$ & $0.245^{+0.000}_{-0.007}$ & 55.4 & 1.1 & R1 \\
 & & & $72.6^{+1.6}_{-0.0}$ & $0^{+43}_{-0}$ & $0.096^{+0.000}_{-0.004}$ & $0.244^{+0.000}_{-0.009}$ & 55.4 & 1.3 & R2 \\
\enddata
\tablenotetext{*}{$W3$ band thermal flux measurements used in the TPM fit are partially saturated.}
\tablecomments{Thermophysical properties of a portion of our final 1847 asteroid sample. Here we limit the sample to the 270 asteroids with the most reliable shape models, as defined in \S\ref{sec.gamma_discuss}. The full table can be found online. The parameters relevant to the TPM inputs as well as the IDs corresponding to the shape models used are given in Table \ref{tab.inputs}. The \citet{Mainzer19} NEATM-derived diameter $D_{\mathrm{WISE}}$ is given where available as a basis for comparison. The volume equivalent diameter $D$, thermal inertia $\Gamma$, Bond albedo $A$, visual geometric albedo $p_\mathrm{V}$, Hapke mean surface slope $\bar{\theta}$, and reduced chi-square $\chi_{\mathrm{red}}^2$ are all outputs of the best-fit TPM solution. While we find the extremely high qualitative roughness of $\bar{\theta} = 55.4\degree$ preferred for many TPM fits, the majority of such cases are subject to partial saturation in the $W3$ band flux measurements, which may be corrupting the TPM fit.}
\end{deluxetable*}
\movetabledown=15mm
\begin{longrotatetable}
\begin{deluxetable*}{rlccccccccccccccccc}
\centering
\tabletypesize{\scriptsize}
\tablecaption{Shape Models and Thermal Infrared Observations\label{tab.inputs}}
\tablehead{
\twocolhead{Asteroid} & \colhead{$\lambda$} & \colhead{$\beta$} & \colhead{$P$} & \colhead{Shape Reference(s)} & \colhead{Mod. ID} & \colhead{Den. LC} & \colhead{Tot. LC} & \colhead{$N_{W3}$} & \colhead{$W3$ Sat.} & \colhead{$N_{W4}$} & \colhead{$W4$ Sat.} & \colhead{$r_{\mathrm{hel}}$} & \colhead{$H$} & \colhead{$G$} & \twocolhead{Taxonomy} & \colhead{Orbit} \\ 
\colhead{} & \colhead{} & \colhead{(deg)} & \colhead{(deg)} & \colhead{(hr)} & \colhead{} & \colhead{} & \colhead{} & \colhead{} & \colhead{} & \colhead{} & \colhead{} & \colhead{} & \colhead{(au)} & \colhead{(mag)} & \colhead{} & \colhead{T1} & \colhead{T2} & \colhead{}
} 
\startdata
5 & Astraea & 126 & 40 & 16.80061 & [1,2,3] & D1 & 7 & 25 & 14 & 0.171 & 14 & 0.000 & 2.896 & 6.8 & 0.15 & S & S & MBA \\
 & & 320 & 45 & 16.80062 & Revised & R1 & 7 & 29 & & & & & & & & & &  \\
 & & 122 & 43 & 16.80061 & Revised & R2 & 7 & 29 & & & & & & & & & &  \\
14 & Irene & 95 & -11 & 15.02986 & [3,4] & D1 & 27 & 33 & 8 & 0.238 & 8 & 0.167 & 2.812 & 6.3 & 0.15 & S & S & MBA \\
 & & 265 & -22 & 15.02990 & Revised & R2 & 27 & 37 & & & & & & & & & &  \\
17 & Thetis & 236 & 19 & 12.26603 & [1,2] & D1 & 14 & 53 & 12 & 0.158 & 12 & 0.000 & 2.799 & 7.8 & 0.15 & S & Sl & MBA \\
 & & 240 & 24 & 12.26603 & Revised & R1 & 14 & 57 & & & & & & & & & &  \\
 & & 57 & 12 & 12.26604 & Revised & R2 & 14 & 57 & & & & & & & & & &  \\
\enddata
\end{deluxetable*}
\end{longrotatetable}
\clearpage

\startlongtable
\begin{deluxetable*}{rlccccccr}
\centering
\tabletypesize{\normalsize}
\tablecaption{Candidates for Yarkovsky Detection\label{tab.yark}}
\tablehead{
	\multicolumn2c{Asteroid} & \colhead{$\Gamma$} & \colhead{$D$} & \colhead{$A$} & \colhead{$\chi^2_{\mathrm{red}}$} & \colhead{$\rho$} & \colhead{Dense LC} & \colhead{$A_2$} \\ 
	\multicolumn2c{} & \colhead{(J\,m$^{-2}$\,s$^{-0.5}$\,K$^{-1}$)} & \colhead{(km)} & \colhead{} & \colhead{} & \colhead{(g cm$^{-3}$)} & \colhead{} & \colhead{(10$^{-15}$ au day$^{-2}$)}
	}
\startdata
1685 & Toro & $210^{+85}_{-55}$ & $3.6^{+0.2}_{-0.1}$ & $0.099^{+0.007}_{-0.009}$ & 1.8 & 2.72\tablenotemark{a} & 67 & $-2.6^{+0.3}_{-0.2}$ \\ 
1865 & Cerberus & $809^{+219}_{-134}$ & $1.2^{+0.0}_{-0.0}$ & $0.061^{+0.000}_{-0.001}$ & 1.6 & 2.72\tablenotemark{a} & 22 & $-7.2^{+0.8}_{-0.6}$ \\ 
2818 & Juvenalis & $57^{+36}_{-57}$ & $4.7^{+0.3}_{-1.0}$ & $0.083^{+0.052}_{-0.010}$ & 1.8 & 2.72\tablenotemark{a} & 0 & $-2.2^{+2.2}_{-0.6}$ \\ 
2991 & Bilbo & $63^{+56}_{-44}$ & $8.4^{+0.9}_{-1.2}$ & $0.031^{+0.012}_{-0.006}$ & 7.5 & 1.38\tablenotemark{b} & 0 & $2.1^{+0.4}_{-0.4}$ \\ 
4928 & Vermeer & $106^{+85}_{-69}$ & $3.7^{+0.4}_{-0.4}$ & $0.085^{+0.025}_{-0.015}$ & 4.0 & 2.71\tablenotemark{b} & 0 & $-2.0^{+0.6}_{-0.4}$ \\ 
5240 & Kwasan & $69^{+206}_{-52}$ & $6.8^{+0.7}_{-0.9}$ & $0.138^{+0.048}_{-0.025}$ & 6.2 & 1.93\tablenotemark{a} & 0 & $-1.9^{+1.0}_{-0.4}$ \\ 
5760 & Mittlefehldt & $61^{+36}_{-27}$ & $9.9^{+0.6}_{-0.8}$ & $0.038^{+0.007}_{-0.004}$ & 2.3 & 1.38\tablenotemark{b} & 0 & $-2.0^{+0.3}_{-0.2}$ \\ 
6581 & Sobers & $55^{+408}_{-53}$ & $4.7^{+0.7}_{-0.5}$ & $0.098^{+0.023}_{-0.023}$ & 3.1 & 2.71\tablenotemark{b} & 0 & $-2.0^{+1.2}_{-0.4}$ \\
\enddata
\tablecomments{Asteroids in our sample with a transverse acceleration parameter $A_2$ greater than the smallest published value of $A_2 = 1.9 \times 10^{-15}$ au day$^{-2}$ for (35107) 1991\,VH. We estimate $A_2$ with the best-fit values for the thermal inertia $\Gamma$, the diameter $D$, and Bond albedo $A$ with Equation \ref{eq.yark}. $\chi^2_{\mathrm{red}}$ refers to that of the TPM fit. The bulk density $\rho$ is either (a) assumed to be equal to the average bulk density of the SMASS II \citep{Bus02} taxonomic type of the asteroid according to \citet{Carry12} or (b) assumed to be equal to the \citet{Krasinsky02} average bulk densities for C-types for $p_V < 0.10$ and S-types for $p_V > 0.10$. Note that a thermal inertia of 0 J m$^{-2}$ s$^{-0.5}$ will drop the $A_2$ parameter to zero, hence the unconstrained lower bound on $A_2$ for 11 of these asteroids. While $A_2$ for (1685) Toro given here is estimated from our derived parameters, we note that it also has a detected Yarkovsky drift of $A_2 = -3.339 \pm 0.6433 \times 10^{-15}$ au day$^{-2}$. Further details can be found in \S\ref{sec.yark}.}
\end{deluxetable*}

\phantom{x}
\end{document}